\documentclass[preprint,iop,amsmath,amssymb]{revtex4-2}
\usepackage{bm}
\usepackage{epsfig}
\usepackage{amsmath}
\usepackage{color}
\newcommand{\nix}[1]{}
\usepackage[unicode=true,colorlinks=true,urlcolor=blue,citecolor=blue]{hyperref}

\begin{document}

\title{
Electric field effect 
on electron gas spins
in two-dimensional magnets 
with strong spin-orbit coupling
}


\author{K.S. Denisov}
\email{denisokonstantin@gmail.com} 
\affiliation{Ioffe Institute, 194021 St.Petersburg, Russia}

\begin{abstract}


The recent rise of material platforms combining magnetism and two-dimensionality of mobile carriers reveals a diverse spectrum of spin-orbit phenomena and stimulates its ongoing theoretical discussions. 
In this work we use the density matrix approach to provide a unified description of subtle microscopic effects governing the electron gas spin behavior in the clean limit upon electric perturbations in two-dimensional magnets with strong spin-orbit coupling. 
We discuss that an inhomogeneity of electrostatic potential generally leads to the electron gas spin tilting with the subsequent formation of equilibrium skyrmion-like spin textures and demonstrate that several microscopic mechanisms of 2DEG spin response are equally important for this effect. 
We analyze the dynamics of 2DEG spin upon an oscillating electric field with a specific focus on the emergent electric dipole spin resonance. 
We address the resonant enhancement of magneto-optical phenomena from the spin precession equation perspective and discuss it in terms of the resonant spin generation. 
We also clarify the connection of both static and dynamic spin phenomena arising in response to a scalar perturbation with the electronic band Berry curvature. 


	
\end{abstract}

\date{\today}

\maketitle

\newpage

\section{Introduction}
\label{s-Intro}

The recent advances in the development of spintronics devices extensively use relativistic spin-orbit properties of free carriers interacting with magnetic layers. 
The spin-orbit coupling (SOC) of charge carriers generally opens up the possibility to deal with the magnetization purely by electrical means; 
the magnetization orientation can be detected electrically by virtue of 
the anisotropic magnetoresistance effect~\cite{gould2004tunneling,moser2007tunneling,Nitin}, 
while electric current-induced spin-orbit torque occurs to be a highly effective tool for switching its direction~\cite{miron2010current,miron2011perpendicular,manchon2019current,song2020spin}. 
Nonstationary dynamics of carriers in presence of SOC
can result in stimulated photon emission, as in case of terahertz spintronic light emitter~\cite{kampfrath2013terahertz,walowski2016perspective,seifert2016efficient,feng2021spintronic} and spin Hall nano-oscillators~\cite{liu2013spectral,awad2017long}.
Apart from kinetic phenomena spin-orbit effects can modify equilibrium spin configurations via indirect RKKY exchange interaction~\cite{kundu2015dzyaloshinskii,ZhuPRL2011,checkelsky2012dirac}
and lead to the formation of magnetic skyrmions~\cite{wiesendanger2016nanoscale,fert2017magnetic} due to Dzyaloshinskii–Moriya terms~\cite{soumyanarayanan2016emergent,moreau2016additive}.   
An efficient charge-to-spin conversion wanted for modern spintronics needs is often realized when turning to a two-dimensional electron gas, as the reduction of the dimensionality tends to be accompanied by the lowering of symmetry and by the subsequent increase in SOC~\cite{soumyanarayanan2016emergent,guillet2021large}.  
There are an increasing number of different material platforms that allow one to combine systematically stronger SOC magnitudes of 2D electrons directly with a magnetic component,  
the examples include 
van der Waals heterostructures~\cite{avsar2020colloquium} 
either proximitized by magnetic layer~\cite{vzutic2019proximitized,xu2018spin,wei2016strong,yang2013proximity,zhao2017enhanced} or being 
intrinsic ferromagnets~\cite{zhong2017van,gong2017discovery,huang2017layer,o2018room}, semiconductor nanostructures doped by magnetic dopants~\cite{lee2000theory,camilleri2001electron}, surface states of magnetic topological insulators~\cite{gong2019experimental,he2017tailoring}, or layered magnetic heterostructures~\cite{rojas2016perpendicular,soumyanarayanan2016emergent}. 
Moreover, combining magnetism with 2D conductive channels additionally offers new functionalities, such as spin tunnel field-effect transistors~\cite{jiang2019spin}, spin inversion effect~\cite{xu2018spin} or novel class of spinterfaces~\cite{dayen2020two}. 


In order to fully benefit from two-dimensional magnetic systems it is of key importance to have a comprehensive understanding of how the spin density of electron gas in a 2D channel responds to an applied electric field, that is the understanding of free electron gas magnetoelectric properties. 
However, a complete microscopic treatment of the related phenomena appears to be extremely challenging, even despite there is a few theoretical approaches effectively dealing with multiband systems (e.g. wave-packet dynamics theory~\cite{sundaram1999wave,culcer2004semiclassical,chang2008berry,xiao2010berry,shindou2005noncommutative}, 
diagrammatic and ab-initio calculations~\cite{yao2004first,zhu2012theory,gradhand2012first,shindou2008gradient}). 
The difficulty lies in the fact that in spin-orbital systems multiple microscopic mechanisms of quite a subtle character often contribute on the equal footing, which hinders a simplified consideration.
In particular, an exchange interaction induced spin splitting in combination with strong spin-orbit coupling generally lead to a geometrical structure of electronic band states featured by nonzero Berry curvature in k-space. 
Treating different spin-related phenomena with account for the electronic band geometry remains an ongoing discussion. 
It covers, for instance, the issues of the Liouville's theorem with account for the Berry phase~\cite{xiao2005berry,bliokh2006hamiltonian,duval2006comment}, 
the Hall conductivity modifications in presence of real-space magnetic textures~\cite{lux2020chiral},
or, concerning the anomalous and spin Hall effects, the interplay between Karplus-Luttinger anomalous velocity and disorder-induced mechanisms~\cite{nagaosa2010ahe,sinitsyn2007anomalous,sinitsyn2007semiclassical,ado2016anomalous}; the latters have recently been enriched by the electron scattering on a pair of impurities~\cite{ado2016anomalous,glazov2020valley}. 
Moreover, when calculating spin-related quantities a specific class of coarse graining effects should be taken into account, as is clearly demonstrated in~\cite{culcer2004semiclassical,xiao2010berry}. 
In this paper we respond to an ever-growing role that two-dimensional magnetic systems plays for spintronics and consider in detail a complex pattern of microscopic effects relevant for the magnetoelectric behavior of 2DEG in the clean limit. 
Based on the density matrix approach
we describe 
the most significant spin-response mechanisms 
of two-dimensional spin-orbital systems 
within the unified  framework, 
reveal the interconnection between  different microscopic effects and clarify its relation to an electronic band geometry.


The theoretical model and the density matrix description are formulated in Sec.~\ref{s-2}. 
In Sec.~\ref{s-static}
we analyze a magnetoelectric effect in thermal equilibrium, namely we consider the formation of equilibrium spin textures and local persistent electric currents arising due to an inhomogeneous electrostatic potential.
We discuss in detail semiclassical electron dynamics with account for a spin-to-momentum locking and identify microscopic mechanisms responsible 
for the magnetoelectric response. 
Namely, we attribute the generation of an extra-spin density directed within 2DEG plane both to the non-adiabatic correction to the electron spin precession and to the correlated change of charge and spin electron densities, the latter scenario is sometimes referred as spin-dipole effect~\cite{culcer2004semiclassical}. 
We provide a unified treatment of these mechanisms using the density matrix,  derive general equations governing the contribution due to each mechanism independently and reveal the role that the Berry curvature plays for the emergent phenomena.


In Sec.~\ref{s-EDSR} 
we turn to the dynamical regime and investigate the 2DEG spin dynamics upon an oscillating electric field. 
We focus specifically on spin resonance phenomena due to electric dipole transitions, also referred as the electric dipole spin resonance (EDSR). 
We derive the precession equation for 2DEG spin density capturing the spin resonance scenario, and clarify the relation of the band states Berry curvature with the spin response susceptibility. 
We also discuss the spin resonance in terms of optical conductivity and 
describe the associated magneto-optical properties of 2DEG. 
In particular, we describe how the EDSR induced generation of the in-plane spin density is accompanied by the resonant enhancement of the Hall conductivity, 
the latter is responsible for magneto-optical Kerr and Faraday effects. 
We classify different spin polarizations emerging in the dynamical regime and present analytic expressions for the spin resonance related optical conductivity. 
\section{Theoretical framework}
\label{s-2}

\subsection{Model band structure}
We consider a two-dimensional electron gas with parabolic bands affected both by the Rashba effect and by an exchange interaction with a magnetic host. 
We assume that the magnetization responsible for the spin splitting is directed along $z$-axis perpendicular to the electron motion plane. 
The so-called Rashba ferromagnet model covers all the physics relevant for our consideration and allows one address the related spin phenomena in the most transparent way. 
The effective Hamiltonian describing this model is given by
\begin{equation}
\label{eq_Ham}
\mathcal{H} = \frac{\bm{k}^2}{2 m}
+ \bm{\Omega}_k \cdot \hat{\bm{S}},  
\end{equation}
here the first term describes the parabolic dispersion with an effective mass $m$, and $\bm{\Omega}_k$ is an effective $k$-space magnetic field acting on the electron spin $\hat{\bm{S}} = \hat{\bm{\sigma}}/2$; $\hat{\bm{\sigma}}$ is the vector of Pauli matrices. 
The field $\bm{\Omega}_k$ leads to a spin splitting of the electronic subbands, in our model $\bm{\Omega}_k$ consists of two parts
\begin{align}
\label{eq_Omegas}
& \bm{\Omega}_k = \bm{\Omega}^{so}(\bm{k}) - \Omega_0 \bm{e}_z, 
\qquad
 \bm{\Omega}^{so}(\bm{k}) = 2 \lambda_{so} \left[ \bm{e}_z \times \bm{k} \right],
\end{align}
where $\bm{\Omega}^{so}(\bm{k})$ describes the spin-orbit Rashba interaction with the coupling constant $\lambda_{so}$, and the second term is due to an exchange interaction with a magnetic background, the parameter $\Omega_0$ describes the corresponding splitting of spin subbands at zero momentum. 
The eigenstates of Eq.~\ref{eq_Ham} Hamiltonian can be written in the following form
\begin{equation}
\label{eq_wave-states}
\psi_k^\pm = e^{i \bm{kr}} |u_k^\pm \rangle, 
\qquad
|u_k^+ \rangle = \frac{1}{\sqrt{2}} \begin{pmatrix}
b_k \\ - i e^{i \varphi } a_k
\end{pmatrix},
\quad
|u_k^- \rangle = \frac{1}{\sqrt{2}} \begin{pmatrix}
-i e^{-i \varphi } a_k  \\ b_k
\end{pmatrix},
\end{equation}
where $(a_k, b_k) = (1 \pm \Omega_0/\Omega_k)^{1/2}$. 
We use the notation $\eta = (\pm)$ for two electron spin subbands. 
The states $\psi_k^\eta$ 
are characterized by the electron spin $\bm{s}_k^\eta = \langle u_k^\eta | \hat{\bm{S}} | u_k^\eta \rangle$ directed either parallel or antiparallel to $\bm{\Omega}_k$
\begin{align}
\label{eq_spin_eq}
& \bm{s}_k^\pm = \pm \frac{1}{2} \bm{n}_k, 
\qquad
\bm{n}_k = \frac{\bm{\Omega}_k}{{\Omega}_k}, 
\qquad
\Omega_k = \sqrt{\Omega_0^2 + (2\lambda_{so} k)^2},
\end{align}
where the unit vector $\bm{n}_k$ points along the direction of $\bm{\Omega}_k$. 

The energy dispersion corresponding to $\eta$-subband is 
$ \varepsilon_k^\eta = k^2/2m + \eta \Omega_k/2$.
The presence of $k$-dependent spin splitting leads to the renormalization of 
effective masses nearby $k\approx 0$, namely $m^\pm = m/(1 \pm \xi)$, where the parameter $\xi \equiv 2 m \lambda_{so}^2/\Omega_0 $. 
We focus on systems with sufficiently strong exchange interaction, when $\Omega_0$ greatly exceeds the spin-orbital coupling. 
We thus take the parameter $\xi < 1$, at that the effective mass $m^- >0$ is positive and the lower energy branch is a monotonic function of the momentum, see Fig.~\ref{f:Q}b. 

Let us discuss the role of the spin splitting terms. 
The presence of the Rashba effect induced spin-momentum locking directly manifests itself in the velocity operator 
\begin{equation}
\label{eq_velocity}
\hat{\bm{v}} = \frac{i}{\hbar} \left[ \mathcal{H}, {\bm{r}} \right] = \frac{\hat{\bm{k}}}{m} + 2 \lambda_{so} [\bm{e}_z \times \hat{\bm{S}}], 
\end{equation}
where the second term is sensitive to the instantaneous direction of the electron spin. While the average velocity for the eigen spin states is determined by the unperturbed spin vector~$\bm{s}_k^\eta$
\begin{equation}
\label{eq_vel0}
\bm{v}_k^\eta \equiv
\langle \psi_k^\eta | \hat{\bm{v}} | \psi_k^\eta \rangle =
 \frac{\bm k}{m} + 2 \lambda_{so} [\bm{e}_z \times \bm{s}_k^\eta], 
\end{equation}
the changes in the direction of an electron spin caused by external fields can directly affect the average of the velocity operator and, correspondingly, influence the orbital motion. 

The presence of a magnetic gap due to the magnetization directed perpendicular to 2DEG plane leads additionally to the fact that electron band states acquire 
a geometric structure. 
Indeed, the electron spin direction in $\bm{k}$-space forms a hedgehog pattern which underlies the appearance of the Berry curvature $\mathcal{F}_k^\eta = i \langle 
\nabla_{\bm k} u_{k}^\eta | \times | \nabla_{\bm k} u_k^\eta
\rangle$. 
For a spin-$1/2$ Hamiltonian this Berry curvature can be expressed as follows
\begin{equation}
\label{eq_BerryField}
\mathcal{F}_k^\eta = \eta\frac{1}{4\pi} \bm{n}_k \cdot \left[
\frac{\partial \bm{n}_k}{\partial {k_x}}
\times 
\frac{\partial \bm{n}_k}{\partial {k_y}}
\right]
= \eta~2 \lambda_{so}^2 \frac{\Omega_0 }{\Omega_k^3},  
\end{equation}
and we keep the notation $\mathcal{F}_k = |\mathcal{F}_k^\eta| $ for its absolute value. 
The total Berry flux $Q_F^\eta$ 
accumulated by electrons from $\eta$ subband 
up to the Fermi energy $\mu$ is given by
\begin{equation}
\label{eq_BerryFlux}
Q_F^\eta = \sum_{k<k_F^\eta} \mathcal{F}_k^\eta = \eta~\frac{1}{4\pi} \left(1 - \frac{\Omega_0}{\Omega_F^\eta} \right), 
\qquad
\Omega_F^\eta = \sqrt{\Omega_0^2 + (2 \lambda_{so} k_F^\eta)^2},
\end{equation}
where $\Omega_F^\pm$ 
is the spin splitting energy for $\eta=(\pm)$ subbands 
at the Fermi energy, see Fig.~\ref{fig:band}b. 
The strong spin-orbit coupling considered in our work means that 
we do not account for the disorder-induced smearing of SOC features of electronic bands.

\subsection{Density matrix approach}

Let us firstly discuss the structure of the density matrix ${f}^0$ for 2DEG in thermal equilibrium without external perturbations. 
The general form is ${f}^0 = (e^{\beta (\hat{\mathcal{H}}-\mu)} + 1 )^{-1}$, 
where $\hat{\mathcal{H}}$ is given by Eq.~\ref{eq_Ham}, 
$\beta$ is the inverse temperature and $\mu$ is the Fermi energy. 
In this work we focus on zero temparature limit $\beta \to \infty$. 
The density matrix $\hat{f}_k^0$ in the momentum representation is a $2\times 2$ matrix which can be presented as follows (we keep hats for spin indices only)
\begin{equation}
	\label{eq_fko}
	\hat{f}_k^0 = \frac{1}{2} n_k^0 + \bm{S}_k^0 \cdot \hat{\bm{\sigma}}. 
\end{equation}
We note that $\hat{f}_k^0$ is diagonal in the basis of eigen states $\psi_k^\pm$, so we can present it as a sum of $\eta = (\pm)$ spin subband contributions  $\hat{f}_k^\eta$
\begin{align}
\label{eq_f_zero}
& \hat{f}_k^0 = 
\hat{f}_k^+ + \hat{f}_k^- , 
\qquad
\hat{f}_k^\eta =   n_k^{\eta}\left(
\frac{1}{2} + \bm{{s}}_k^{\eta} \cdot \hat{\bm{\sigma}} 
\right), 
\end{align}
where 
$n_k^\eta = (e^{\beta (\varepsilon_k^\eta - \mu)}+1)^{-1}$ is the Fermi-Dirac distribution function of electrons in the spin subband with energy $\varepsilon_k^\eta$. 
The terms in Eq.~\ref{eq_fko} are given $n_k^0 = n_k^+ + n_k^-$, and $\bm{S}_k^0 = n_k^+ \bm{s}_k^+ + n_k^- \bm{s}_k^-$, here 
$\bm{s}_k^\eta$ corresponds to the 
eigen spin states from Eq.~\ref{eq_spin_eq}. 
The equilibrium spin density $\bm{S}_0$ is directed perpendicular to the 2DEG plane
\begin{equation}
\label{eq_S-equil}
\bm{S}_0 = \frac{1}{2}
\sum_k {\rm Sp}\left( \hat{f}_k^0 \cdot \hat{\bm{\sigma}} \right) = 
\sum_{k} \left(n_k^+ \bm{s}_k^+ +  n_k^- \bm{s}_k^-\right) = \bm{e}_z~\frac{\Omega_F^- - \Omega_F^+}{16 \pi \lambda_{so}^2}. 
\end{equation}
We note that when both spin subbands are populated 
($\mu > \Omega_0/2$) the equilibrium spin density takes value 
${S}_0 = m \Omega_0/4\pi$ 
independent of the Fermi energy, this is specific for Hamiltonian from Eq.~\ref{eq_Ham}. 

The application of a scalar potential $U(\bm{r},t)$ deviates the electron distribution from Eq.\ref{eq_f_zero}. 
In this paper we focus on spatially smooth perturbations ($ k_F  \cdot \nabla_k \ll 1$ and $\lambda_F \cdot \nabla_r \ll 1$) and 
study the electron gas response in the classical limit. 
For this purpose we introduce the Wigner density matrix $\hat{f}_k(\bm{r},t)$ in the following form
\begin{align}
\label{eq_f-9}
& \hat{f}_k(\bm{r},t) = \frac{1}{2} n_k(\bm{r},t) + \bm{S}_k(\bm{r},t) \cdot \hat{\bm \sigma},
\end{align}
where $n_k(\bm{r},t), \bm{S}_k(\bm{r},t)$ 
can be treated as particle and spin distribution functions locally in real space. In particular, the 2DEG spin density perturbation emerging in the real space at point $\bm{r}$ can be found from
\begin{equation}
\label{eq_spin-pertrt-den}
\delta \bm{S}(\bm{r},t) = \frac{1}{2} \sum_k {\rm Sp}\left( \hat{f}_k(\bm{r},t) \cdot \hat{\bm{\sigma}} \right) - \bm{S}_0 = \sum_k  \bm{S}_k(\bm{r},t) - \bm{S}_0.
\end{equation}
In the clean limit 
$\hat{f}_k(\bm{r},t)$ satisfies the kinetic equation~\cite{Dyakonov} 
\begin{align}
\label{eq_kin}
& \frac{\partial \hat{f}_k}{\partial t} + \frac{1}{2} 
\Bigl\{
\left(\hat{\bm{v}} \cdot \nabla_{\bm{r}}\right) ; \hat{f}_k
\Bigr\}
- \left[ \bm{\Omega}_k \times \bm{S}_k\right] \cdot \hat{\bm{\sigma}}
+ \left(\bm{F} \cdot \nabla_{\bm{k}}\right) \hat{f}_k = 0, 
\end{align}
where $\{;\}$ stands for the anticommutator,  $\nabla_{\bm{r},\bm{k}}$ are the nabla operators, 
$\bm{F}(\bm{r},t) = - \nabla_{\bm{r}} U(\bm{r},t)$ describes the dynamical force acting on electrons, and the third term takes into account the precession of the electron spin in the effective magnetic field $\bm{\Omega}_k$. 
Let us draw the attention to the anticommutator type of ordering  between $\hat{\bm{v}}$ and $\hat{f}_k$ that appears in the second term. 
This ordering directly stems from the Wigner transformation procedure~\cite{mishchenko2004spin} and it is especially important to 
describe accurately the response in the inhomogeneous regime.  

\section{Static spin textures}
\label{s-static}

\begin{figure}[t]
	\centering
	\includegraphics[width=0.5\textwidth]{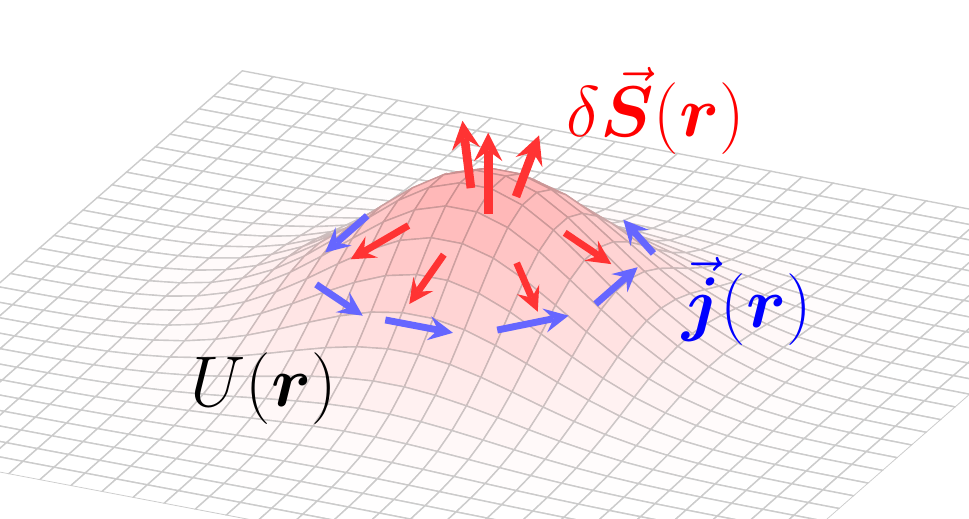}
	\caption{Formation of skyrmion-like spin textures and the distribution of the persistent electric currents nearby electrostatic defects.} 
	\label{fig:skyrm}
\end{figure}

We start our analysis by inspecting the redistribution of the 2DEG charge and spin densities nearby smooth electrostatic defects, such as Coulomb centres or gating potential perturbations. 
The geometric character of electronic band states and the associated nonzero Berry curvature underline the appearance of chiral spin textures and adjoint persistent electric currents that surround electrostatic potential inhomogeneity, see Fig.~\ref{fig:skyrm}. 
In~\cite{denisov2019chiral} 
we used the Kubo formalism to address the nonlocal regime of the spin density response due to short-range impurities. 
In this section, instead, we provide a detailed semiclassical description of this phenomenon and accompany it by the comprehensive physical analysis. 

\subsection{General mechanisms of the intrinsic spin generation}

\label{s_3a}

Let us qualitatively discuss the effect of the electron spin non-adiabatic rotation upon the precession in a slowly varying magnetic field~\cite{sinitsyn2005disorder,AharonovStern,sinova2004universal}.
We start by considering the precession equation for an electron spin $\bm{s}$ rotating upon a time-dependent  frequency $\bm{\Omega}(t)$
\begin{equation}
\label{eq_prec}
\frac{d \bm{s}}{dt} = \left[\bm{\Omega}(t) \times \bm{s} \right]. 
\end{equation}
Assuming the adiabatically slow rotation of $\bm{\Omega}(t)$, i.e. that the characteristic time $\tau$ of its variation satisfies $\Omega \tau \gg 1$, 
the zero-order solution of the precession equation simply describes the electron spin $\bm{s}^0(t) = \bm{\Omega}(t)/2|\bm{\Omega}(t)|$ remaining co-aligned with the instant direction of $\bm{\Omega}(t)$. 
However, the adiabatic rotation of $\bm{s}^0(t)$ can be maintained only due to the appearance of the non-adiabatic correction $\delta \bm{s}(t)$ directed perpendicular to the instant vector $\bm{\Omega}(t)$. 
Naturally, this correction exists in the first order in $(\Omega \tau)^{-1}$ and it can be found from the precession equation keeping only the leading term due to $\bm{s}^0(t)$ in the time derivative 
\begin{equation}
\label{eq_s-prec1}
\frac{d \bm{s}^0}{dt} = \left[\bm{\Omega}(t) \times \delta \bm{s}(t) \right] 
\quad
\rightarrow
\quad \delta {\bm{s}}(t) = \frac{1}{2 \Omega^3} \left[\bm{\Omega} \times \frac{d\bm{\Omega}}{dt}\right]. 
\end{equation}
The appearance of $\delta {s} \propto  (\Omega \tau)^{-1} {s}^0$ is a general property of the precession equation. 
Naturally, this is also valid when a Larmor frequency stems from an effective magnetic field in k-space due to a spin-orbit coupling. 
In this case, however, 
the vector $\bm{\Omega}_k$ that governs the spin dynamics of 
an electron with momentum $\bm{k}$ varies in time only provided that the electron momentum does not remain constant along its trajectory $\dot{\bm{k}} \neq 0$, which is the case if $\bm{F} \neq 0$. 
The non-adiabatic spin component acquired by an electron can be estimated from Eq.~\ref{eq_s-prec1} by replacing the time derivative by $d/dt \to \dot{\bm{k}}\cdot \nabla_{\bm{k}}$
\begin{equation}
\label{eq_s-prec2}
(\dot{\bm{k}}\cdot \nabla_{\bm{k}} )\bm{s}_k^0 = \left[\bm{\Omega}_k \times \delta \bm{s}_k
\right] 
\quad
\rightarrow
\quad
\delta \bm{s}_k = \frac{1}{2\Omega_k^3} \left[
\bm{\Omega}_k \times 
(\dot{\bm{k}}\cdot \nabla_{\bm{k}})
\bm{\Omega}_k
\right]. 
\end{equation}
We conclude that an electron moving along its classical trajectory with finite acceleration has its spin always slightly tilted compared to the instantaneous direction of $\bm{\Omega}_k$. Moreover, in view of the spin-momentum locking such an intrinsically generated extra-spin leads to the change in the 
electron velocity $\delta \bm{v}_k = 2\lambda_{so} (\bm{e}_z \times \delta \bm{s}_k)$. 


The second spin-related phenomenon being important for the collective response of 2DEG concerns the spin-dipole effect~\cite{culcer2004semiclassical}. 
This mechanism is relevant when the single electron density $|\psi(\bm{r})|^2$ deviates from the homogeneous distribution and acquires some finite $\bm{r}$-dependence nearby an inhomogeneity. Let us consider an electron at the unperturbed plane-wave state $\psi_k^\pm$ from Eq.~\ref{eq_wave-states} with the momentum $\bm{k}$, its spin $\bm{s}_k^\pm$ is determined by $\bm{\Omega} _k$. 
The corresponding density $|\psi_k^\pm|^2$ is spatially homogeneous. 
In fact, the smooth spatial variation of the density for such electron is possible only provided that its wave-function gets an admixture of other plane-wave band states $\psi_{k'}^\pm$ with momenta $\bm k'$ slighty differing from $\bm{k}$. 
Essentially, the added states have different spin orientation $\bm s_{k'}^\pm \neq \bm{s}_k^\pm$, so the resulting average spin density appears to be slightly tilted. In terms of the wave-packet dynamics~\cite{culcer2006geometrical,culcer2004semiclassical} the mixing of spin-orbital states leads to the fact that the charge and spin centers of the electron wave-packet do not coincide, which creates an additional spin polarization. 
This scenario is specifically important for localized electron states~\cite{kavokin2008spin,denisov2018hall}. 
We emphasize that the spin-dipole effect is essentially connected with the spatial variation of the electron density. In particular, if a given external field keeps an electron gas in the homogeneous state, the spin-dipole contribution will be absent. 
The appearance of the non-adiabatic correction from Eq.~\ref{eq_s-prec2}, on the contrary, is not connected with the change of an electron density, it simply tracks the exact electron spin dynamics along quasiclassical trajectories. 


\subsection{Density matrix in a static inhomogeneous setting}

We proceed with giving a rigorous description of the outlined phenomena based on the kinetic equation for the density matrix. 
Let us consider an electron gas subjected to an electrostatic potential $U(\bm{r})$ smoothly varying in space. 
Since the unperturbed density matrix $\hat{f}_k^0 = \hat{f}_k^+ + \hat{f}_k^-$ given by Eq.~\ref{eq_f_zero} 
has two parts 
corresponding to $\eta = (\pm)$ subband states, 
the linear response correction 
$\delta \hat{f}_k(\bm{r}) = \hat{f}_k(\bm{r}) - \hat{f}_k^0$ 
will be determined independently by two subband terms 
$ \delta \hat{f}_k(\bm{r}) =  \delta \hat{f}_k^+(\bm{r}) +  \delta \hat{f}_k^-(\bm{r})$.  
We present the corresponding correction $ \delta \hat{f}_k^\eta $ 
as follows
\begin{align}
&
\delta \hat{f}_k^\eta(\bm r) = \frac{1}{2} \delta n_k^\eta(\bm{r}) + \delta \bm{S}_k^\eta(\bm{r}) \cdot \hat{\bm{\sigma}}, 
\end{align}
where $\delta n_k^\eta(\bm{r}), \delta \bm{S}_k^\eta(\bm{r}) $ 
are the perturbations of the electron density and spin distribution functions, respectively. 

The key suggestion implemented in this paper is to use the following ansats for the linear response spin density 
\begin{equation}
\label{eq_devSpin}
\delta \bm{S}_k^\eta(\bm{r}) = \delta n_k^\eta(\bm{r})  \bm{s}_k^\eta + 
n_k^\eta \delta \bm{s}_k^\eta(\bm{r}) + \delta \bm{\mathcal{S}}_k^\eta(\bm{r}), 
\end{equation}
where we took into account all possible types of $\delta \bm{S}_k^\eta(\bm{r})$ variation.  
Indeed, the first term describes the change of the electron spin distribution due to the change in the density $\delta n_k^\eta$. 
The second term corresponds to the change of the spin vector $\delta \bm{s}_k^\eta$ for each individual electron independently of the electron number distribution. 
The third term is the remaining linear-order variation, which is essentially neither due to $\delta n_k^\eta(\bm{r})$ or $\delta \bm{s}_k^\eta(\bm{r})$ separately; thus $\delta \bm{\mathcal{S}}_k^\eta$ describes the correlated change of both the electron spin and charge densities. 
Naturally, the second and the third terms in this expansion turn out to describe the non-adiabatic spin tilting and the spin-dipole effects, respectively. 

We proceed with calculating $\delta \bm{S}_k^\eta(\bm{r})$ from the kinetic equation~\ref{eq_kin}. 
In what follows we keep in Eq.~\ref{eq_kin} only the terms linear in $U$ and $\bm{F}=-\nabla_{\bm{r}} U$. 
In this limit the change of the electron density $\delta n_k^\eta$ can be  determined independently from the scalar part of Eq.~\ref{eq_kin}. 
Taking the trace over Eq.~\ref{eq_kin} we get 
\begin{equation}
\label{eq_den}
\left(\bm{v}_k^{\eta} \cdot \nabla_r 
+ \bm{F}(\bm{r}) \cdot \nabla_k 
\right) n_k^\eta(\bm{r})
=0. 
\end{equation}
Here $\bm{v}_k^\eta$ is the electron group velocity given by Eq.~\ref{eq_vel0}. 
In the linear response regime the correction $\delta n_k^\eta$ is given by: $\delta n_k^\eta(\bm{r}) = U(\bm{r}) (\partial n_k^\eta/\partial \varepsilon)$, where $\varepsilon$ is the electron energy. 
The change in the overall 2DEG density is
$\delta n(\bm{r}) = \delta n^+(\bm{r})  + \delta n^-(\bm{r})$, where $\delta n^\eta(\bm{r}) = -\nu_F^\eta~ U(\bm{r})$ and 
$\nu_F^\eta$ is the density of states in $\eta$ subbands 
taken at the Fermi energy. 
Correspondingly, the perturbation of the spin density Eq.~\ref{eq_spin-pertrt-den} due to the first term in Eq.~\ref{eq_devSpin} is given by
\begin{equation}
\label{eq_delta-S1}
\delta \bm{S}^{(1)}(\bm r) = \sum_{k,\eta} \bm{s}_k^\eta \cdot \delta n_k^\eta(\bm{r})
= \bm{e}_z \Omega_0 \left( \frac{\nu_F^+}{\Omega_{F}^+} - \frac{\nu_F^-}{\Omega_{F}^-} \right) U(\bm{r}).
\end{equation}

The term $\delta \bm{S}^{(1)}(\bm r)$ is responsible for the change in the
out-of-plane spin density component and it appears even if there is no spin-orbit interaction. 
A complex spin-orbital electron dynamics is responsible for an extra spin response described by $\delta{\bm{s}}_k^\eta$ and $\delta \bm{\mathcal{S}}_k^\eta$. 
We notice that $\delta \bm{s}_k^\eta, \delta \bm{\mathcal{S}}_k^\eta$ are absent in a homogeneous setting, thus the expansion of $\delta \bm{s}_k^\eta, \delta \bm{\mathcal{S}}_k^\eta$ starts with the linear term $\nabla_r U$. 
Taking the trace over Eq.~\ref{eq_kin} multiplied by $\hat{\bm{\sigma}}$ and keeping only the terms linear in $\nabla_{\bm{r}}$ gradient we get
\begin{align}
& 
\left[
\bm{\Omega}_k \times \delta \bm{s}_k^\eta(\bm{r})
\right]  -
\left(\bm{F}(\bm{r}) \cdot \nabla_k \right)
\bm{s}_k^\eta
 =0,
\label{eq_s-nonad}
\\
& 
\left[
\bm{\Omega}_k \times \delta \bm{\mathcal{S}}_k^\eta(\bm{r})
\right]
+ 
\left[
\bm{s}_k^\eta \times 
\left(
\bm{s}_k^\eta
\times 
{\bm{\Omega}^{so}({\nabla}_{\bm r} n_k^\eta)}
\right)
\right] =0, 
\label{eq_s-correl}
\end{align}
where $\bm{\Omega}^{so}(\nabla_r n_k^\eta)$ is obtained from Eq.~\ref{eq_Omegas} by replacing $\bm{k} \to \nabla_{\bm r} n_k^\eta(\bm{r})$. 

Let us comment on the relation between  $\delta \bm{s}_k^\eta, \delta \bm{\mathcal{S}}_k^\eta$ 
and the previously described kinematic effects. 
The first equation Eq.~\ref{eq_s-nonad}
can be satisfied by changing the electron spin vector $\delta \bm{s}_k^\eta$ independently of a particular density distribution $n_k^\eta$, 
it thus indeed describes the spin rotation of individual electrons due to the precession in the effective magnetic field~$\bm{\Omega}_k$. 
Naturally, the nonzero term $\delta \bm{s}_k^\eta$ is exactly the non-adiabatic correction to the instant spin vector $\bm{s}_k^\eta$ which follows adiabatically the local direction of  $\bm{\Omega}_k$.  
The solution of the equation~\ref{eq_s-nonad}
replicates the result from Eq.\ref{eq_s-prec2}
\begin{equation}
\label{eq_s-prec-new}
\delta \bm{s}_k^\eta(\bm{r}) =
\eta \frac{1}{2\Omega_k^3} \Bigl[
\bm{\Omega}_k \times 
(\bm{F}(\bm{r}) \cdot \nabla_{\bm{k}})
\bm{\Omega}_k
\Bigr]. 
\end{equation}
It is worth noting that  $\delta \bm{s}_k^\eta$ is nonlinear with respect to 
$\bm{\Omega}_k$. 
The second equation Eq.~\ref{eq_s-correl} describes the appearance of 
$\delta \bm{\mathcal{S}}_k^\eta$, 
the general form of the solution is given by
\begin{equation}
\label{eq_spin-dip}
\delta \bm{\mathcal{S}}_k^\eta(\bm{r}) = - \frac{1}{4 \Omega_k^2} \left[
\bm{\Omega}_k 
\times 
\bm{\Omega}^{so}({\nabla}_r n_k^\eta)
\right].
\end{equation}
Importantly, the additional spin density $\delta \bm{\mathcal{S}}_k^\eta$ responds directly to the spatial gradient of the electron density $\nabla_r n_k^\eta(\bm{r})$ entering in $\bm{\Omega}^{so}$. 
In fact, this allows us to refer 
$\delta \bm{\mathcal{S}}_k^\eta$ as the correlational term: it is neither due to the independent change in the number of electrons or due to the individual electron spin rotation. Instead, $\delta \bm{\mathcal{S}}_k^\eta$ describes the simultaneous change in the electron spin due to the variation in its spatial density, it is indeed relevant to the spin-dipole effect.

\subsection{Interplay between microscopic mechanisms and the role of Berry curvature}

The explicit evaluation of extra-spin density terms from Eq.~\ref{eq_s-prec-new},\ref{eq_spin-dip} for the Rashba ferromagnet model gives the following expressions
\begin{align}
\label{eq_static_sko}
&\delta \bm{s}_k^\eta = \eta \frac{e \mathcal{F}_k}{2 \lambda_{so}} \cdot \bm{E}(\bm{r}) - \eta \frac{2e\lambda_{so}^2}{\Omega_k^3} \left[\bm{k}\times \bm{E}(\bm{r})\right], 
\\
\label{eq_static_sko-2}
& \delta \bm{\mathcal{S}}_k^\eta = - \mathcal{F}_k\cdot \frac{\Omega_k}{4 \lambda_{so}} \nabla_{\bm{r}} n_k^\eta(\bm{r}) + \eta \frac{\lambda_{so}}{2 \Omega_k^2}  \bm{e}_z \left(
\bm{\Omega}_k \cdot \nabla_{\bm{r}} 
\right) n_k^\eta(\bm{r}), 
\end{align}
where $\mathcal{F}_k$ is the magnitude 
of the Berry curvature from Eq.~\ref{eq_BerryField}, 
and the density gradient $\nabla_r n_k^\eta(\bm{r}) = - e\bm{E}(\bm{r}) (\partial n_k^\eta/\partial \varepsilon)$ is due to the redistribution of electrons in the vicinity of an electrostatic potential inhomogeneity. 

We note that various terms from Eqs.~\ref{eq_static_sko},~\ref{eq_static_sko-2}
give rise to quite different spin phenomena. 
For instance, the second terms in $\delta \bm{s}_k^\eta, \delta \bm{\mathcal{S}}_k^\eta$ depend on the electron momentum direction and they are particularly important for the generation of spin currents in nonmagnetic systems (they survive at $\Omega_0 \to 0$); the second term in $\delta \bm{s}_k^\eta$ is responsible for the universal spin Hall conductivity~\cite{sinova2004universal}. 
Alternatively, it keeps significance for spin dynamics, see the details in Sec.~\ref{s-EDSR}. 
Below we focus on the local magnetoelectric effect, that is the appearance of an equilibrium  spin density in response to the local electric field. 
This phenomenon stems from the first terms in  $\delta \bm{s}_k^\eta, \delta \bm{\mathcal{S}}_k^\eta$; they can directly generate an additional spin density at a given point in a space as they survive averaging over the electron momentum direction. 
Moreover, 
these terms can be explicitly expressed in terms of the Berry curvature, thus they are specific for topological systems. 

\begin{figure}[t]
	\centering
	\includegraphics[width=0.5\textwidth]{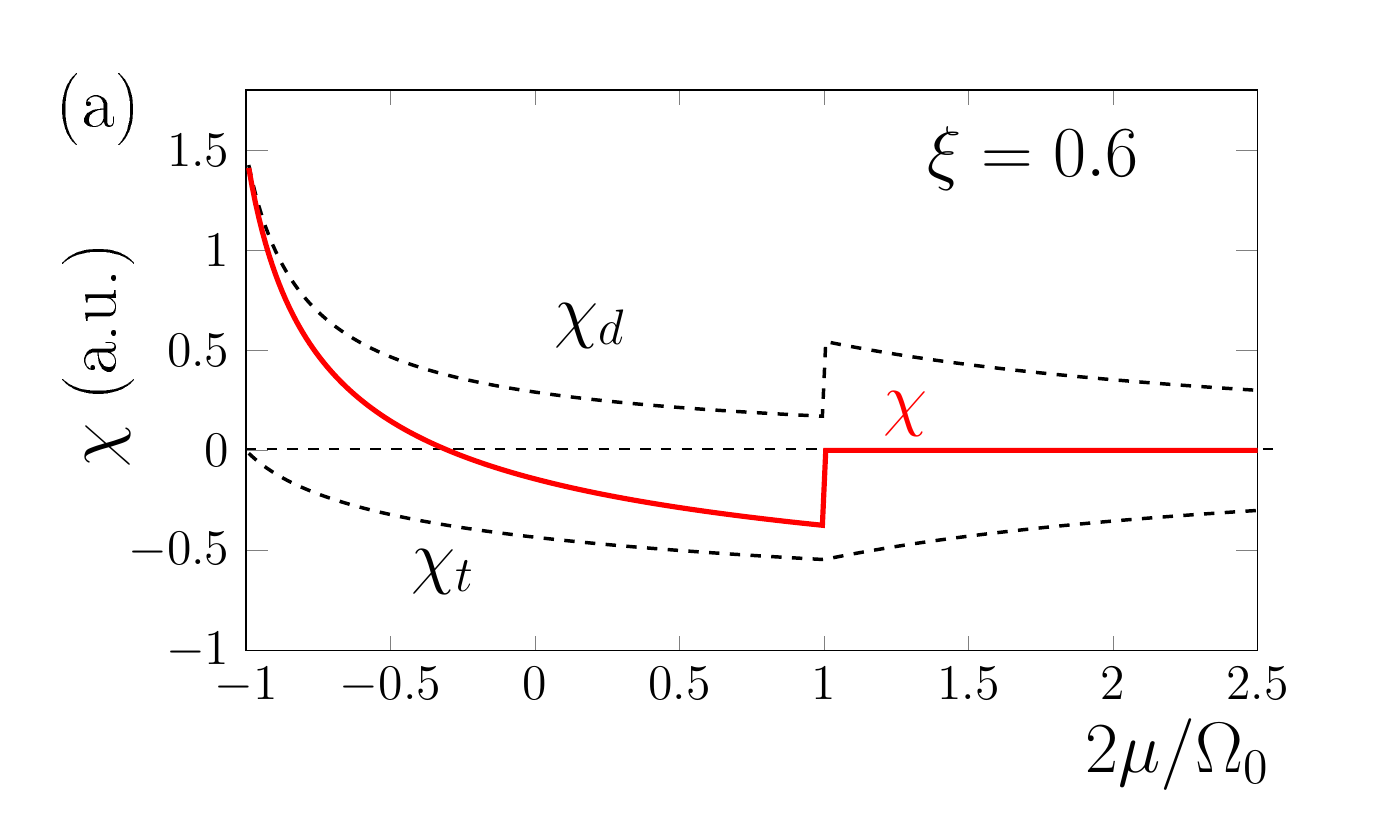}%
	\includegraphics[width=0.5\textwidth]{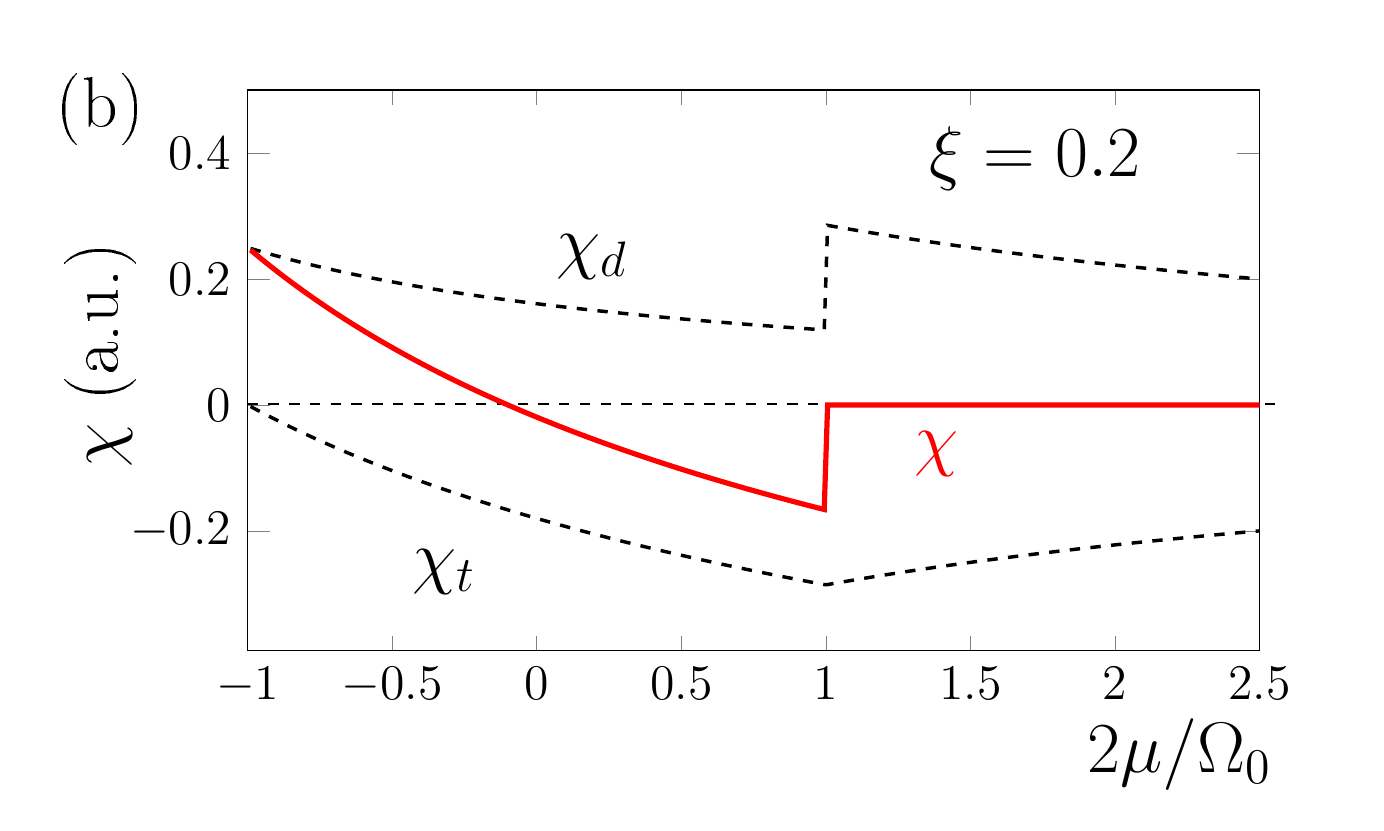}%
	\caption{
		The dependence of the susceptibility $\chi=\chi_t + \chi_{d}$ on the Fermi energy for two values of $\xi$ parameter: (a) $\xi = 0.6$ and (b) $\xi = 0.2$.} 
	\label{fR:06}
\end{figure}

The equilibrium spin density perturbations coupled with the Berry curvature of electronic states have only in-plane components; 
substituting 
Eqs.~\ref{eq_static_sko},~\ref{eq_static_sko-2} to the spin density perturbation from Eq.~\ref{eq_spin-pertrt-den} we get
\begin{align}
& \delta \bm{S}_\parallel(\bm{r}) = \sum_{k, \eta} 
n_k^\eta \delta \bm{s}_k^\eta(\bm{r}) + \delta \bm{\mathcal{S}}_k^\eta(\bm{r}) \equiv \left(
\chi_{t} + \chi_{d}
\right) \cdot \bm{E}(\bm{r}), 
\end{align}
where the magnetoelectric susceptibilities  $\chi_{t,d}$ correspond to  
the non-adiabatic spin tilting and spin-dipole effects, respectively. 
The evaluated expressions for $\chi_t,\chi_{d}$ are given by
\begin{equation}
\label{eq_chi1d}
 \chi_t =  \frac{e}{2\lambda_{so}} \left(Q_F^+ + Q_F^- \right), 
 \qquad
 \chi_{d} = - e\frac{\lambda_{so} \Omega_0}{2} \left(\frac{\nu_F^+}{\Omega_{F+}^2} + \frac{\nu_F^-}{\Omega_{F-}^2} \right), 
\end{equation}
where $Q_F^\pm$ is the total Berry flux from Eq.~\ref{eq_BerryFlux}. 
It is important to emphasize that both the non-adiabatic spin tilting and the spin-dipole effects are equally important to describe correctly the emergent spin patterns in 2DEG. 
In Fig.~\ref{fR:06} we plot the dependence of the overall spin-response coefficient $\chi \equiv \chi_t + \chi_{d}$ (solid lines) along with the partial contributions from $\chi_t$ and $\chi_{d}$ (dotted lines) on the electron gas Fermi energy $\mu$. 
We note that the terms $\chi_t$ and $\chi_{d}$ are generally of the same order of magnitude. Moreover, in case when the electron gas populates both spin subbands $\mu > \Omega_0/2$ the overall response entirely disappears $\chi_t + \chi_{d}=0$ (this feature was previously noted by~\cite{denisov2019chiral,mishchenko2014equilibrium}). 
In the opposite case when electrons fill only the lowest spin-subband $\mu<\Omega_0/2$ the terms $\chi_t,\chi_{d}$ have opposite signs, which results in the sign-altering dependence of $\chi$ on the Fermi energy. 
We finally note that when either the spin-orbit coupling or the exchange interaction is absent, the coefficients $\chi_t=\chi_{d}=0$ turn to zero and the corresponding equilibrium spin patterns disappear. 




\subsection{Discussion}

Let us discuss the physical significance of the described phenomena. 
We firstly  comment on the role that intrinsic mechanisms described by Eqs.~\ref{eq_static_sko},~\ref{eq_static_sko-2} 
play for the charge and spin transport 
on distances that greatly exceed the mean free path. 
The non-adiabatic spin precession lies in the basis of 
the Karplus-Luttinger mechanism of the anomalous Hall effect (AHE)~\cite{karplus1954hall,jungwirth2002anomalous,niu1985quantized}, of
the so-called intrinsic mechanisms of the spin Hall (SHE)~\cite{sinova2004universal} 
and spin-galvanic effects~\cite{culcer2007generation}. 
However, in order to estimate correctly the overall electron gas response one has to additionally examine the disorder effects. 
In particular, the intrinsic 
contribution to AHE, which is due to the anomalous velocity term  $\delta \bm{v}_k^\eta \propto e \mathcal{F}_{k}^\eta \cdot \left[\bm{e}_z \times \bm{E}\right]$, 
is generally cancelled out by the contributions due to side-jump scattering  processes~\cite{Dyakonov,sinitsyn2007anomalous,glazov2020valley}.
Alternatively, considering the generation of spin currents upon the applied homogeneous electric field one has to carefully account for the emergent nonequilibrium phenomena~\cite{khaetskii2006nonexistence,culcer2007generation,chen2014spin,wang2012diffusive}; e.g. 
the spin Hall current due to the intrinsic  mechanism is often compensated by the nonequilibrium spin current arising nearby the sample boundaries~\cite{mishchenko2004spin,Erratum:Nomura,raimondi2005spin}.

However, the contributions  $\delta \bm{s}_k^\eta, \delta \bm{\mathcal{S}}_k^\eta$ preserve the importance in the nondissipative regime, when the underlying electrostatic perturbation varies at the distances much smaller than the mean free path. 
In particular, this matters for 2DEG charge and spin distribution around an ionized impurity, at that the typical spatial scale under consideration is the Thomas-Fermi screening length. 
The distribution of an excessive 2DEG spin density 
emerging around an axially symmetric perturbation 
forms a skyrmion-like vortex pattern which 
is schematically shown in Fig.~\ref{fig:skyrm}. 
One concludes that a smooth electrostatic potential disorder in topological spin polarized 2DEG inevitably generates chiral spin textures, which can be particularly important for the transport properties of the corresponding system; 
the formation of non-collinear spin order generally leads to the topological Hall effect~\cite{denisov2018general,ishizuka2018spin,taguchi2003magnetic}. 
Moreover, in view of the spin-velocity coupling 
the formation of a mesoscopic in-plane spin density is accompanied by the generation of the persistent electrical current density 
$\bm{j}(\bm{r}) = e~2 \lambda_{so} \left[\bm{e}_z \times \delta {\bm{S}(\bm r)}\right]$. 
In this regard an axially symmetric perturbation from Fig.~\ref{fig:skyrm} is additionally featured by radially propagating electric currents. 
The presence of local equilibrium currents also maintains the orbital magnetization, this effect has been considered in~\cite{mishchenko2014equilibrium}.

It is worth mentioning that the considered magnetoelectric susceptibility of free electrons generally opens up a possibility to directly affect the host magnetization by a mesoscopic electric perturbation. 
The electric field-induced 2DEG spin density lies in 2D channel plane and it is perpendicular to the orientation of host magnetization, thus it is able to produce torque-like effects. However, these issues remain poorly investigated, even despite its importance for the magnetization control at nanoscales. 







The microscopic mechanisms under consideration are general for multiband systems.
In the appendix~\ref{ap-1} we present the connection of our method with the wave-packet quasiclassical technique used in~\cite{sundaram1999wave,culcer2004semiclassical,chang2008berry}. 
In the appendix~\ref{ap-2} we relate $\delta \bm{s}_k^\eta, \delta \bm{\mathcal{S}}_k^\eta$ 
to the Kubo formula method for the charge-spin correlation functions used in~\cite{denisov2019chiral}. 
In particular, 
we show that the non-adiabatic spin precession is described by the interband correlation functions, while the spin-dipole effect stems from the intraband ones. 
\section{Spin dynamics and magneto-optical effects}
\label{s-EDSR}


\subsection{Electric dipole spin resonance}

\label{s:4a}

In this section we focus on the electron gas spin dynamics in presence of an oscillating electric field and describe the corresponding optical properties of a magnetic two-dimensional system. 
The optical response of a 2D conductive channel is generally encoded in the optical conductivity $\sigma(\omega)$. 
In particular,  the absorption coefficient $\alpha(\omega) = (4\pi/c) {\rm Re}[\sigma_{xx}(\omega)] $ is connected with the longitudinal part of conductivity $\sigma_{xx}$. 
Also, since the time-reversal symmetry is broken in presence of magnetism, different magneto-optical effects are possible, e.g. the magneto-optical Kerr effect (MOKE), that is the rotation of the reflected light polarization by the complex Kerr angle $\phi_K$. 
MOKE generally appears in a conductive media due to nonzero optical Hall conductivity $\sigma_{H}(\omega)$, for a 2D layer and normal incidence~\cite{yang2020magneto} one can expess $\phi_K = \sigma_H / \sigma_{xx} \sqrt{1+(4\pi i /\omega) \sigma_{xx}} $. 
Importantly, the considered geometry 
opens up the possibility to realize the resonant enhancement of the Hall conductivity and, thus, of the related magneto-optical effects. 

Commonly, MOKE is seen to acquire a resonance structure due to 
interband transitions affected by the combined effect of the spin-orbit coupling and the electron spin polarization; 
the corresponding intrinsic contributions to the Hall conductivity at finite frequencies have been investigated in a number of papers~\cite{mainkar1996first,guo1995band,yao2004first,uba1996optical}.   
The general idea that we are going to explore in this paper and which stands in the basis for the enhancement of magneto-optical phenomena is that the optical properties of magnetic 2D systems can be understood in terms of the electric dipole spin resonance~(EDSR). 
Correspondingly, the part of the optical conductivity responsible for the resonant features can be directly related to the resonantly generated spin density of 2DEG.

\begin{figure}[t]
	\centering
	\includegraphics[width=0.6\textwidth]{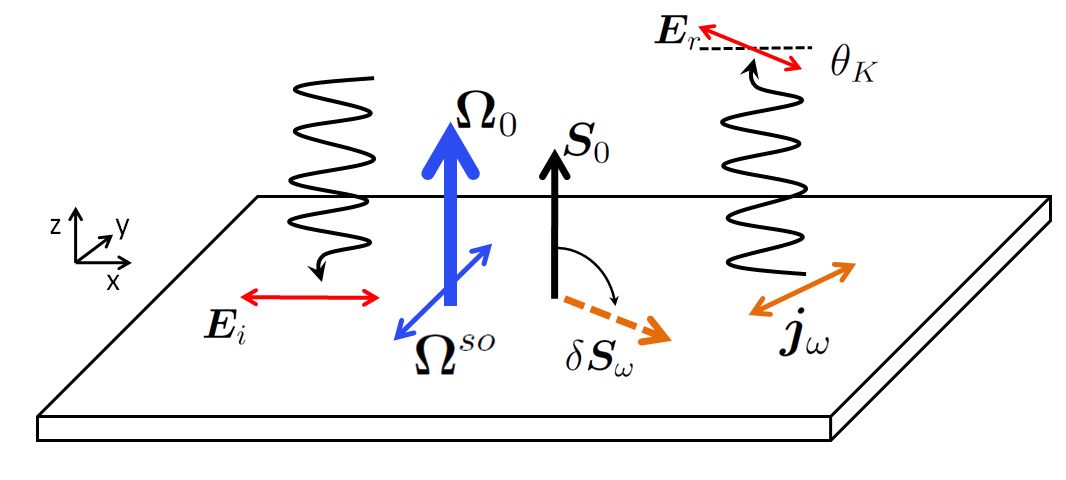}
	\caption{The electric dipole spin resonance scheme and the appearance of MOKE due to the resonant Hall current generation $\bm{j}_\omega \propto \left[\bm{e}_z \times \delta \bm{S}_\omega \right]$ . 
	} 
	\label{fig:band}
\end{figure}

Let us illustrate this process in more detail, see Fig.~\ref{fig:band}. 
The exchange interaction field gives rise to a momentum-independent Zeeman splitting of the electron spin subbands, for the considered geometry it is directed perpendicular to the 2DEG plane. 
In fact, the spin-orbit interaction can be viewed as  $\bm{k}$-dependent effective magnetic field $\bm{\Omega}^{so}(\bm{k})$ acting on electron spins. 
The applied in-plane ac-electric field $\bm{E}_{\omega}e^{-i\omega t}$ causes the electron's momentum oscillations $\delta \bm{k} \propto \bm{E}_{\omega}e^{-i\omega t}$, so the associated spin-orbital field also oscillates with frequency $\omega$. 
We note that $\bm{\Omega}^{so}(\bm{k})$ is perpendicular to the out-of-plane exchange interaction component $\bm{\Omega}_0$.
Naturally, this makes it possible to induce spin transitions when the electric field frequency coincides with the magnitude of the Zeeman spin splitting $\hbar \omega = \Omega_0$, which is exactly the EDSR scheme~\cite{rashba1960properties}. 
This spin resonance causes the equilibrium electron spin density $\bm{S}_0 \parallel \bm{e}_z$ from Eq.~\ref{eq_S-equil} to rotate onto 2DEG plane, 
thus resonantly generating an excessive in-plane spin density $\delta \bm{S}_\omega$. 
In view of the spin-orbit coupling Eq.~\ref{eq_velocity} between the velocity and spin operators, the accumulation of $\delta \bm{S}_\omega$ immediately  leads to a resonant enhancement of the associated electric current density 
$\delta \bm{j}_\omega = 2e ~\lambda_{so} \left[\bm{e}_z \times \delta \bm{S}_\omega \right] $ and of the corresponding contribution to the optical conductivity. 
Importantly, the in-plane spin density appears in tilted polarization with respect to the vector of the electric field, see Fig~\ref{fig:band}. 
In particular, the manifestation of the nonzero Berry curvature lies in the fact, that 
there exists the "perpendicular" polarization of the spin density, which gives rise to the anomalous velocity $\delta \bm{v}_k^{\eta} \propto e \mathcal{F}_k^\eta \cdot \left[\bm{e}_z \times \bm{E}\right]$ 
directed perpendicular to $\bm{E}_\omega$ and responsible for the  
the magneto-optical effects.
The resonant generation of the spin density in this polarization leads to the enhancement of $\sigma_{H}(\omega)$.

\subsection{Density matrix in the dynamical regime}



Let us consider an oscillating electric field $\bm{E}_\omega e^{-i\omega t}$ applied in plane of the electron gas. 
We assume that the system remains homogeneous and present $\hat{f}_k$ 
in the following form
\begin{equation}
\hat{f}_k(t) = \frac{1}{2} n_k(t) 
+ \bm{S}_k(t) \cdot \hat{\bm{\sigma}}. 
\end{equation}
We keep to the high-frequency regime when $\omega$ greatly exceeds the typical inverse relaxation time $\tau_{sc}^{-1}$ due to the scattering processes. 
The distribution function $n_k(t) = n_k + \delta n_k(\omega) e^{-i\omega t}$ satisfies the scalar part of the kinetic equation Eq.~\ref{eq_kin}
\begin{equation}
\label{eq_dens}
\frac{\partial n_k(t)}{\partial t} - e\left(\bm{E}(t)\cdot \nabla_k\right) n_k(t) =0, 
\end{equation}
Since the equilibrium part contains terms from both spin subbands $n_k = n_k^+ + n_k^-$, the linear response perturbation $\delta n_k(\omega) = \delta n_k^+(\omega) + \delta n_k^-(\omega)$ 
generally contains two contributions 
\begin{equation}
	\label{eq_delta_nk}
\delta n_k^\eta(\omega) = 
- \frac{e \bm{E}\cdot \bm{v}_k^{\eta}}{i \omega}  \left(- \frac{\partial n_k^\eta}{\partial \varepsilon}\right). 
\end{equation}
The equation governing 2DEG spin dynamics is obtained similarly to Eq.~\ref{eq_kin} and reads as
\begin{align}
& \frac{\partial \bm{S}_k(t)}{\partial t} - 
\left[\bm{\Omega}_k \times \bm{S}_{k}(t) \right] +  e\left(\bm{E}(t)\cdot \nabla_k\right) \bm{S}_k(t) =0. 
\end{align}
At zero electric field 
this equation describes the electron spin precession around $\bm{\Omega}_k$. 
The static regime solution in this case corresponds to the equilibrium spin distribution $\bm{S}_k^\pm \parallel \bm{\Omega}_k$ directed parallel or antiparallel to the spin splitting field, while 
the non-stationary solution describes the electron spin precession around $\bm{\Omega}_k$ with an eigenfrequency $\Omega_k$. 
The nonzero $\bm{E}$, in its turn, drives the spin dynamics due to the spin transfer in the momentum space. 
Naturally, when the frequency of an external field $\omega$ coincides with the precession frequency of the $\bm{k}$-electrons,  the EDSR conditions are fulfilled leading to the resonant rotation. 
This rotation occurs with the Rabi frequency $\omega_R \propto \lambda_{so} E$, which goes to zero at small electric fields. 
Naturally, in case of vanishing $\omega_R$ we can consider the linear response regime with $\bm{S}_k(t) = \bm{S}_k^0 + \delta \bm{S}_k(\omega) e^{-i\omega t}$ differing from the equilibrium value $ \bm{S}_k^0 = n_k^+ \bm{s}_k^+ + n_k^- \bm{s}_k^-$ by the linear-order correction $\delta \bm{S}_k(\omega)$. This is justified when 
the ongoing evolution of $\bm{S}_k(t)$ due to the Rabi oscillations is interrupted by the spin relaxation processes. We thus introduce the phenomenological spin relaxation rate $\Gamma$ and assume $ \omega_R \ll \Gamma \ll \Omega_0$. 


In the linear response regime we can consider the spin response 
$\delta \bm{S}_k(\omega) = \delta \bm{S}_k^+(\omega) + \delta \bm{S}_k^-(\omega) $ 
independently for each spin subband (recall that $ \bm{S}_k^0 = n_k^+ \bm{s}_k^+ + n_k^- \bm{s}_k^-$). 
It is convenient to present the linearized part in the following way 
$\delta \bm{S}_k^\eta(\omega) = \delta n_k^\eta(\omega) \bm{s}_k^\eta + n_k^\eta \delta \bm{s}_k^\eta(\omega)$, 
where 
$\delta n_k^\eta(\omega)$ is determined by Eq.~\ref{eq_delta_nk}
and the equation for $\delta \bm{s}_k^\eta(\omega)$ is given by
\begin{equation}
\left(- i \omega + \Gamma \right) \delta \bm{s}_k^\eta(\omega) -
\left[\bm{\Omega}_k \times \delta \bm{s}_{k}^\eta(\omega) \right] +  e\left(\bm{E}_\omega\cdot \nabla_k\right) \bm{s}_k^\eta =0. 
\end{equation}
Let us introduce the notation $\delta \bm{s}_{k0}^\eta \equiv \delta \bm{s}_k^\eta(\omega \to 0)$ for the additional electron spin density from 
Eq.~\ref{eq_s-prec-new} emerging in the static limit, we note that $(\delta \bm{s}_{k0}^\eta \cdot \bm{\Omega}_k) =0$. 
The third term in this equation can be presented as follows
$
 e\left(\bm{E}_\omega\cdot \nabla_k\right) \bm{s}_k^\eta = \left[\bm{\Omega}_k \times \delta \bm{s}_{k0}^\eta  \right]
$.
The spin density perturbation $\delta \bm{s}_k^\eta(\omega)$ lies in the plane perpendicular to $\bm{\Omega}_k$, the two independent polarizations for $\delta \bm{s}_k^\eta(\omega)$ are given by $\delta \bm{s}_{k0}^\eta$ and $\left[\bm{n}_k \times \delta \bm{s}_{k0}^\eta \right]$, 
where $\bm{n}_k = \bm{\Omega}_k/\Omega_k$ from Eq.~\ref{eq_spin_eq}.  
The solution of the precession equation can be written in terms of these two vectors as follows
\begin{equation}
\label{eq_s-dynam}
\delta \bm{s}_k^\eta(\omega) = - \frac{\Omega_k^2}{\left( \omega - \Omega_k + i \Gamma \right) 
	\left( \omega + \Omega_k + i \Gamma \right)
}
\left(
\delta \bm{s}_{k0}^\eta + \frac{-i \omega + \Gamma}{\Omega_k} 
\left[\bm{n}_k \times \delta \bm{s}_{k0}^\eta \right]
\right). 
\end{equation}
The first term is directly due to the finite-frequency evolution of the non-adiabatic spin tilt mechanism. 
The second term exists only at finite frequencies and it arises from the electron spin retardation in the momentum space. 
The denominator has a pole structure which reflects the EDSR 
with the multiple resonances determined by $\omega = \Omega_k$. 

The resulting correction to the density matrix can be presented as a sum of two terms $	\delta \hat{f}_k = e^{-i\omega t}( \delta \hat{f}_k^{\rm den} + \delta \hat{f}_k^{\rm spin})$, where $\delta \hat{f}_k^{\rm den, spin}$ 
take the following form
\begin{align}
&	\delta \hat{f}_k^{\rm den} = 
	  \frac{1}{2} \left( \delta n_k^+(\omega) + \delta n_k^-(\omega)  \right)  
+  \left( \delta n_k^+(\omega)  \bm{s}_k^+  + \delta n_k^-(\omega)  \bm{s}_k^-\right)
 \cdot \hat{\bm{\sigma}},
\\
& 	\delta \hat{f}_k^{\rm spin} = 
	\left(n_k^+ \delta \bm{s}_k^+(\omega) + n_k^- \delta \bm{s}_k^-(\omega) \right)\cdot \hat{\bm{\sigma}}.
\end{align}


\subsection{Resonant spin response and optical conductivity}

We start the discussion of the optical conductivity. 
The contribution $\delta \hat{f}_k^{\rm den}$  is related specifically to the perturbation of the electron density and it gives rise to the dominant part of the longitudinal conductivity
\begin{equation}
	\label{eq_long_sigma}
\bm{j}_\omega = e
	\sum_{k, \eta} \delta n_k^\eta(\omega) \cdot \bm{v}_k^\eta
	= \sigma_{xx}^0(\omega) \bm{E}_\omega, 
	\qquad
\sigma_{xx}^0(\omega)	=  \frac{i e^2}{\omega}~\frac{
		{v_{F+}^2 \nu_F^+} + {v_{F-}^2 \nu_F^-}
	}{2}. 
\end{equation}
This is simply the Drude conductivity at finite frequency and it describes nondissipative retardation of the 2DEG density in ac-electric field. 
On the contrary, the term 
$\delta \hat{f}_k^{\rm spin}$ 
is due to the spin rotation only. 
This contribution is responsible for the spin resonance related phenomena and below we consider its role in more detail. 

The density of an electric current $\delta \bm{j}_\omega$ 
emerging due to the spin part of the density matrix $\delta \hat{f}_k^{\rm spin}$ is coupled with an induced in-plane spin density $\delta \bm{S}_\omega $ of 2DEG 
\begin{align}
&	\delta \bm{j}_\omega = 2e \cdot \lambda_{so} \left[\bm{e}_z \times \delta \bm{S}_\omega \right], 
\\
& \delta \bm{S}_\omega  = \frac{1}{2}
\sum_k {\rm Sp}\left( \delta \hat{f}_k^{\rm spin} \cdot \hat{\bm{\sigma}} \right) =
\sum_{k, \eta} n_k^\eta ~ \delta \bm{s}_k^\eta(\omega). 
\end{align}
Since $\delta \bm{s}_k^\eta(\omega)$ generally has two polarizations, see Eq.~\ref{eq_s-dynam}, the overall spin $\delta \bm{S}_\omega $ and correspondingly the associated current $\delta \bm{j}_\omega$ are also featured by two independent~polarizations
\begin{align}
& \delta \bm{S}_\omega = 
 \chi_l (\omega) \left[
\bm{e}_z \times \bm{E}_\omega
\right] + \chi_H(\omega) \bm{E}_\omega,
\\
& 
\delta \bm{j}_\omega = 
\sigma_{l}(\omega)
\bm{E}_\omega
+ 
\sigma_H(\omega)
\left[\bm{e}_z \times \bm{E}_\omega \right],
\end{align}
where 
$\sigma_{l,H}(\omega) = 2 e\lambda_{so}\chi_{l,H}(\omega)$. 
By this we identified the contributions to the optical conductivity related to the magnetoelectric spin susceptibility. 

The correction to the longitudinal conductivity $\sigma_l(\omega)$ is related to the retardation term $\left[\bm{n}_k \times \delta \bm{s}_{k0}^\eta \right]$ in Eq.~\ref{eq_s-dynam}. 
Using the formula Eq.~\ref{eq_static_sko} for $\delta \bm{s}_{k0}^\eta$ and 
averaging over momentum directions we get 
(below we restore the Planck constant $\hbar$)
\begin{equation}
\label{eq_sigmal}	
	\sigma_{l}(\omega) =  -i e^2 \cdot 
	\sum_k \frac{\left( n_k^- - n_k^+ \right) \hbar \omega}{\left(\hbar \omega - \Omega_k + i \Gamma \right) 
		\left(\hbar \omega + \Omega_k + i \Gamma \right)} \frac{\lambda_{so}^2}{\Omega_k} \left(1 + \frac{\Omega_0^2}{\Omega_k^2}\right). 
\end{equation}
The straightforward calculation of this integral gives
\begin{equation}
	\sigma_l(\omega) = - \frac{i e^2}{16 \pi \hbar} 
	\left[
	\frac{2 \Omega_0^2}{\hbar \omega} \left( \frac{1}{\Omega_{\rm min}} - \frac{1}{\Omega_F^{-}}  \right)
	+ \left(
	1 + \frac{\Omega_0^2}{(\hbar \omega)^2}
	\right)
 \ln{
	\left(
	\frac{\hbar \omega + \Omega_F^-}{\hbar \omega  - \Omega_F^-} \cdot \frac{\hbar  \omega -\Omega_{\rm min}}{\hbar \omega + \Omega_{\rm min}}
	\right)
}
	\right], 
\end{equation}
where $\Omega_{\rm min} = \Omega_0$ for $\mu < \Omega_0/2$ and 
$\Omega_{\rm min} = \Omega_F^+$ for $\mu > \Omega_0/2$. 
The expression from above remains well-defined at $\Gamma \to 0$. 
In fact, the poles $\hbar \omega = \Omega_k$ in the denominator of $\delta \bm{s}_k^\eta(\omega)$ lie in the continuum spectrum, so the overall response of closely lying resonances merges onto the $\omega$-regular curve featured by the Van Hove singularities at the edges of the spin splittins $\hbar \omega = (\Omega_0, \Omega_F^\pm)$. 

\begin{figure}[t]
	\centering
	\includegraphics[width=0.55\textwidth]{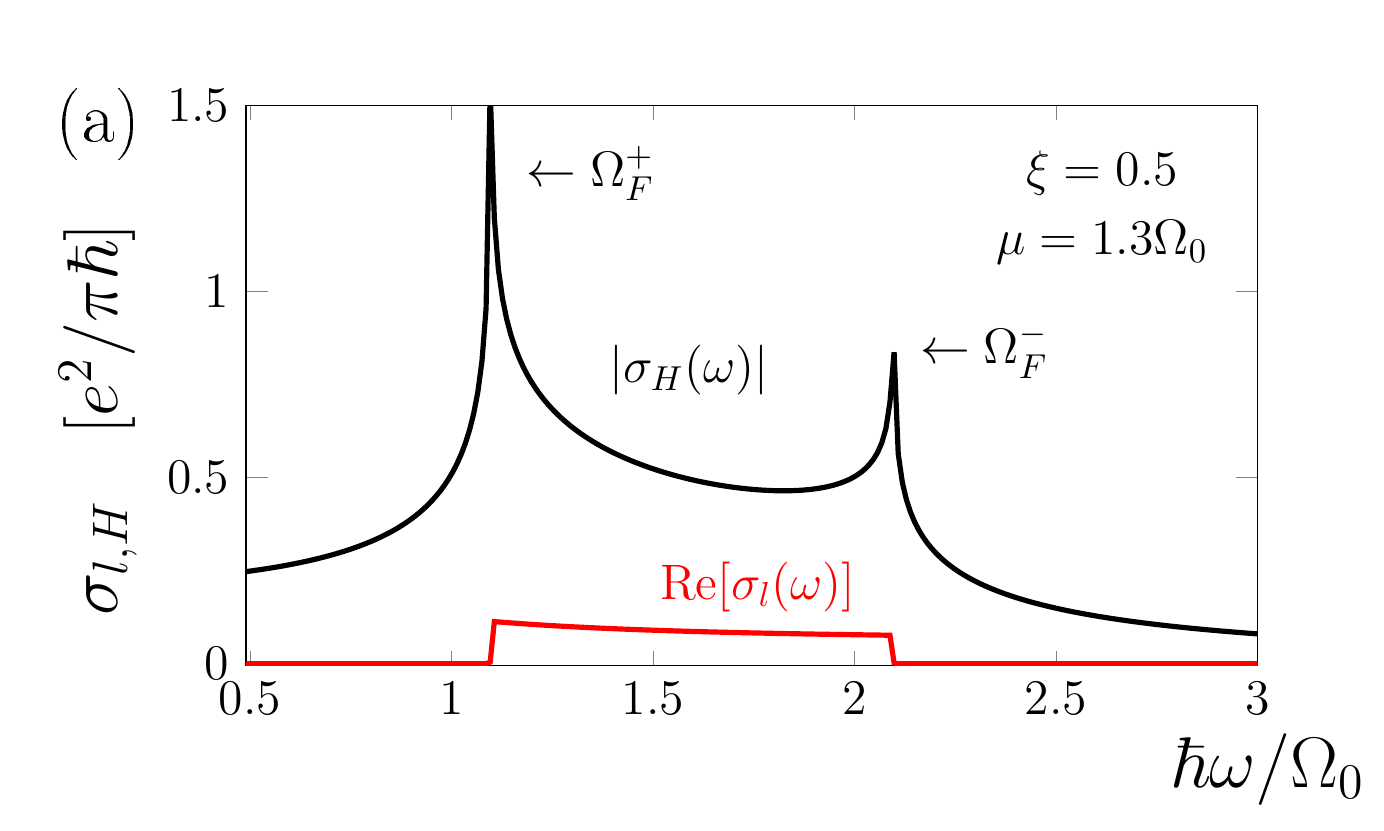}
	\qquad
	\includegraphics[width=0.35\textwidth]{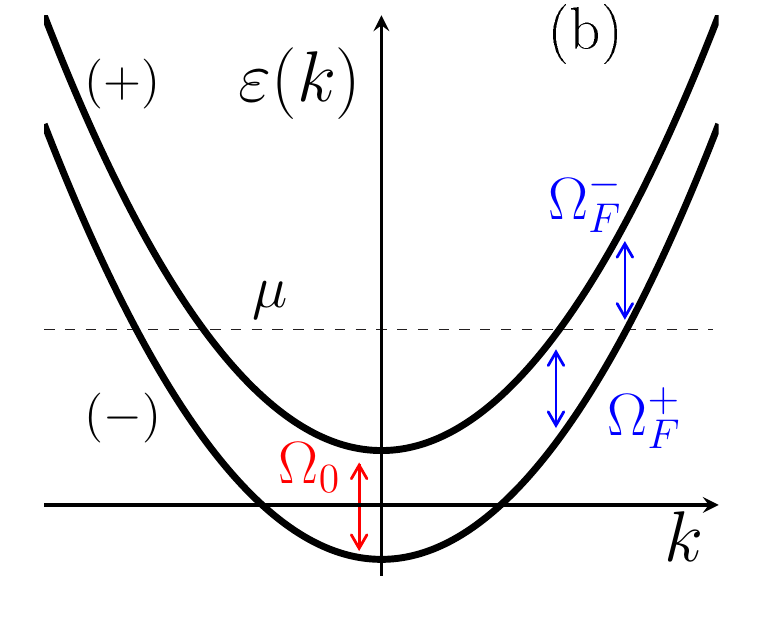}
	\caption{
		(a) The dependence of optical conductivities $\sigma_{l,H}$ on frequency exhibits a resonant structure due to EDSR.
		(b) Electron band structure and the transition energies $\Omega_{F}^{\pm}$ at the Fermi~level.} 
	\label{f:Q}
\end{figure}

The real part of the longitudinal conductivity describes the energy dissipation.
The presence of the resonant poles in Eq.~\ref{eq_s-dynam} reflects the appearance of a finite absorption. 
Indeed, the absorption coefficient is nonzero in the frequency range $\Omega_{\rm min} < \hbar \omega < \Omega_F^-$ (see Fig.~\ref{f:Q}b) corresponding to EDSR, the expression is given by
\begin{equation}
	\label{eq_alpha_abs}
	\alpha(\omega) = \frac{4\pi}{c} {\rm Re}\left[\sigma_l(\omega)\right]
	= \frac{\pi e^2}{4 \hbar c} \left[1 + \left(\frac{\Omega_0}{\hbar \omega}\right)^2 \right],
	\qquad
	\Omega_{\rm min} < \hbar \omega < \Omega_F^-.
\end{equation}

\begin{figure}[t]
	\centering
	\includegraphics[width=0.5\textwidth]{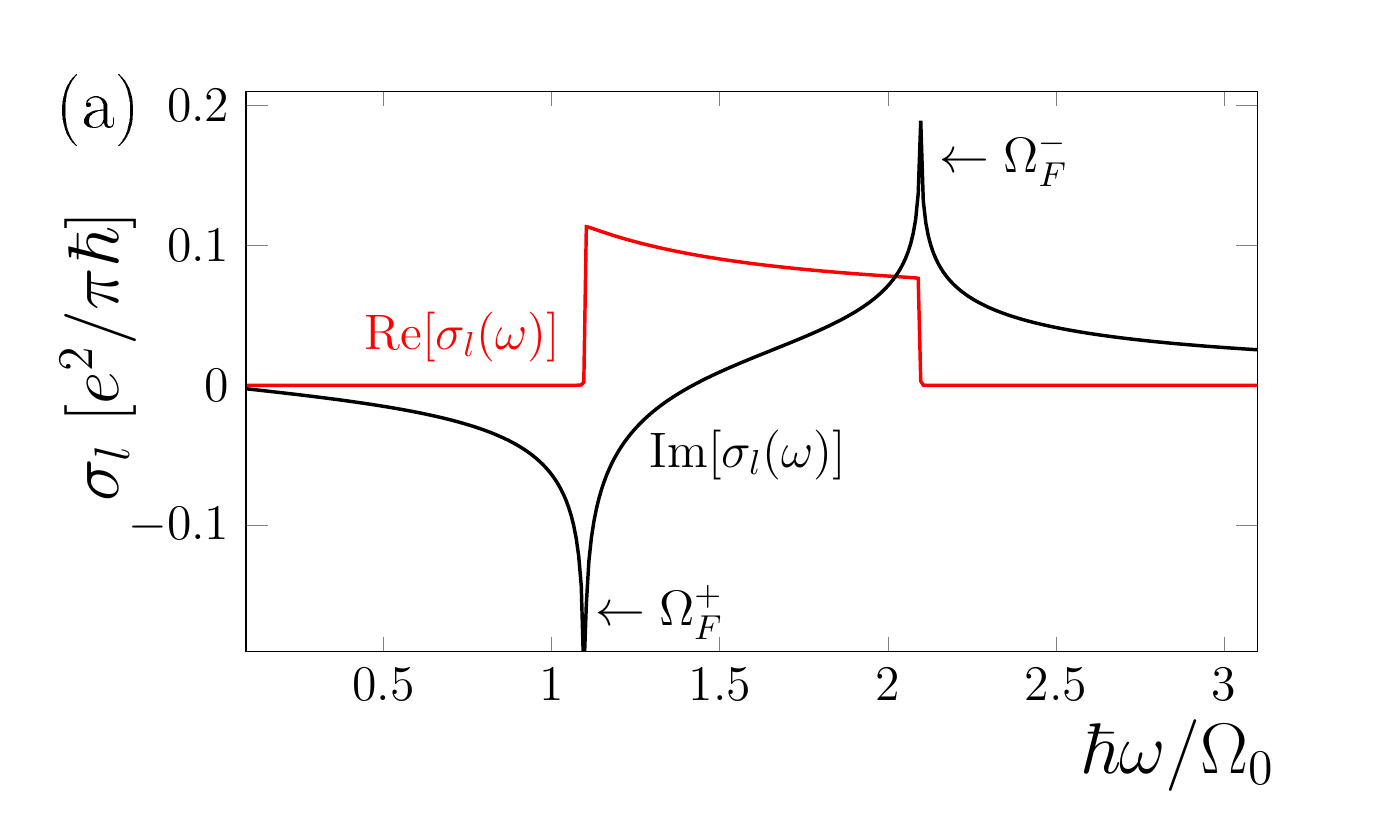}
	\includegraphics[width=0.5\textwidth]{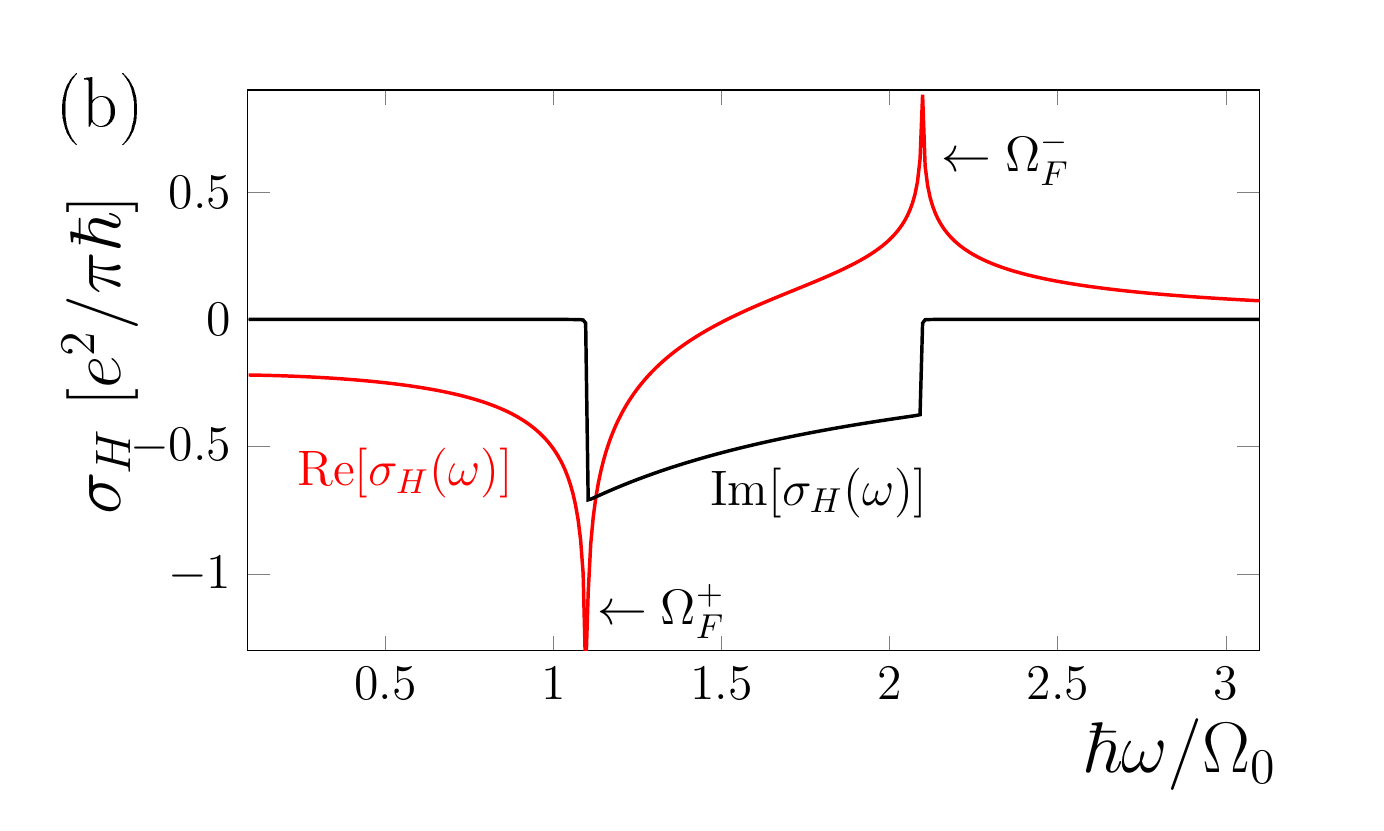}
	\caption{
		The dependence of the optical conductivities $\sigma_{l}(\omega)$ (frame a) and 
		$\sigma_{H}(\omega)$ (frame b)
		on the frequency $\omega$, the parameters $\xi = 0.5, \mu = 1.3~\Omega_0$.} 
	\label{f:sigma}
\end{figure}

The Hall conductivity $\sigma_H(\omega)$ stems from the Berry curvature related term in $\delta \bm{s}_{k0}^\eta$. Taking into account Eqs.~\ref{eq_static_sko},~\ref{eq_s-dynam} and averaging over the momentum direction we express $\sigma_H(\omega)$ 
\begin{equation}
\label{eq_chi1_omega}
\sigma_H(\omega) = 
- \frac{e^2}{\hbar} \sum_{k}  \frac{ (n_k^- - n_k^+)~\Omega_k^2}{\left( \hbar\omega - \Omega_k + i \Gamma \right) 
	\left( \hbar \omega + \Omega_k + i \Gamma \right)
} \cdot \mathcal{F}_k. 
\end{equation}
The evaluation of this expression gives the following result
\begin{align}
\label{eq_chi1-resonant}
& 
\sigma_{H}(\omega) = - \frac{e^2}{4 \pi \hbar} \frac{\Omega_0}{\hbar \omega}  \ln{
	\left(
	\frac{\hbar \omega + \Omega_F^-}{\hbar \omega  - \Omega_F^-} \cdot \frac{\hbar  \omega -\Omega_{\rm min}}{\hbar \omega + \Omega_{\rm min}}
\right)
}.
\end{align}
Importantly, the Hall conductivity has the same resonance-aware logarithmic term as $\sigma_l(\omega)$. 
Fig.~\ref{f:Q} demonstrates the resonant enhancement of the Hall conductivity 
in the EDSR absorption frequency range. Namely, we plot the dependence of
${\rm Re}[\sigma_l(\omega)]$ and the absolute value $|\sigma_H(\omega)|$ 
on the electric field frequency. It is clearly seen from Fig.~\ref{f:Q} that the increase in $|\sigma_H(\omega)|$ magnitude occurs exactly in the same frequency range where ${\rm Re}[\sigma_l(\omega)] \neq 0$ is nonzero.
In Fig.~\ref{f:sigma} we plot the dependences of real and imaginary parts of the spin-resonance related optical conductivities $\sigma_{l,H}(\omega)$ on frequency. The parameters are the same as in Fig.~\ref{f:Q}. The Van Hove singularities give rise to the pronounced peaks in $|\sigma_{l,H}(\omega)|$ 
at the boundary of the absorption band $ \hbar \omega = \Omega_{F}^+, \Omega_F^-$. 
For the parameters taken in this plot ($\mu = 1.3 \Omega_0$) the lower boundary is determined by $\Omega_{F}^+$, see Fig.~\ref{f:Q}, as the electrons populate both spin subbands. We also note that the behavior of $\sigma_{l,H}(\omega)$ 
when approaching the static limit $\omega \to 0$ is different, see Fig.~\ref{f:sigma}.  
While the longitudinal part goes to zero $\sigma_{l} \to 0$, the Hall conductivity has a finite nonzero limit $\sigma_{H} \to (e^2/\hbar) (Q_F^+ + Q_F^-) $ determined by the total Berry flux $Q_F^\pm$ from Eq.~\ref{eq_BerryFlux} and reflecting the appearance of persistent electric currents associated with the magnetoelectric susceptibility. 
In the static limit, however, the accurate calculation of $\sigma_{H}$ for a macroscopic sample requires 
one to take into account the disorder effect~\cite{AHE-Sinova}. 


\subsection{Discussion}
\label{s_Disc}

The calculations of the optical conductivity of multiband systems is typically performed using the Kubo formula~\cite{mainkar1996first,guo1995band,yao2004first,uba1996optical}. 
In the Appendix~\ref{ap-3} we relate the spin polarization and the density contributions from  the density matrix approach with different terms from the Kubo formalism. In Table~\ref{table:1} we summarize the correspondence between these approaches; naturally the spin resonance related terms are connected with the interband contributions $\sigma^{\rm inter}$ to the conductivity.  

Let us comment on the role of spin relaxation and electron scattering. The multiple-peak structure of $\sigma_{l,H}(\omega)$ visible in Fig.~\ref{f:Q} can be well resolved only provided that the spin-orbit interaction splitting ($|\Omega_F^\pm - \Omega_0 | \gg \tau_{sc}^{-1}$) exceeds the energy broadening due to scattering processes. This requires rather strong spin-orbit coupling. In the opposite case, 
the resonance profile of $\sigma_{l,H}(\omega)$ 
will merge onto the single resonant-peak structure centered at $\Omega_0$ with the line-shape sensitive to particular scattering and spin relaxation processes, in analogy with EDSR due to an electron gas in nonmagnetic semiconductors~\cite{duckheim2006electric}. 
Interestingly, the Hall conductivity can possess an additional information on spin relaxation times. 

\begin{table}[]
	\begin{tabular}{|c|c|c|c|}
		\hline
		Kubo formula \quad &  \quad$\sigma_{xx}^{\rm intra} $ \quad &  \quad$\sigma_{xx}^{\rm inter} $ \quad   &  \quad$\sigma_{H}^{\rm inter} $ \quad   \\
		\hline
		Density matrix	\quad &\quad $ \displaystyle \delta \hat{f}_k^{\rm den} $ \quad & \quad $ \displaystyle  n_k^\eta 
		\left[\bm{n}_p \times \delta \bm{s}_{k0}^\eta \right] \cdot \hat{\bm{\sigma}}    $ \quad   & \quad $ \displaystyle n_k^\eta  \delta \bm{s}_{k0}^\eta \cdot \hat{\bm{\sigma}}    $ \quad    \\
		\hline
	\end{tabular}
	\caption{Density matrix and Kubo formula correspondence}
	\label{table:1}
\end{table}

We note that the finite absorption due to the electric dipole spin resonance in 2DEG is not strictly limited to the case when the Zeeman field has an out-of-plane component. In fact, most of the EDSR experiments with 2DEG in nonmagnetic semiconductors~\cite{kato2004coherent,duckheim2007resonant,schulte2005electron} 
were carried out for the in-plane magnetic field geometry. 
This is particularly useful when one aims to suppress the orbital quantization effects and to focus on the spin-related response only. 
On the contrary, combining spin-orbital electronic channels with magnetism allows one to orient the Zeeman field perpendicular to the 2DEG plane without breaking the spectrum onto Landau levels.
Moreover, in this setting the electron band states are featured by the appearance of a topological structure. 
Studying experimentally the electronic spin resonance phenomena in these systems seems of high interest as EDSR has an extra degree of freedom that is the strong enhancement of the adjoint magneto-optical effects. 

Finally, the presented interpretation of the magneto-optical effects enhancement in terms of spin resonance is equally relevant for other two-dimensional models beyond Rashba ferromagnets. For instance, e.g. massive Dirac metals~\cite{catarina2020magneto}, honeycomb lattices~\cite{shah2020magneto} or Haldane model~\cite{pratama2020circular} demonstrate similar resonant features of the Hall conductivity.

\section*{Summary}

In summary,  we have considered various spin-orbital phenomena leading to a nontrivial behavior of an electron gas spin density upon application of the electric field in two-dimensional magnets. Based on the density matrix formalism we identified different microscopic mechanisms responsible for the 2DEG spin tilting in presence of an inhomogeneous electrostatic potential, and described microscopic features of spin resonance upon oscillating electric field with specific focus on optical conductivity and magneto-optical phenomena. We traced the connection of the considered spin phenomena with the Berry curvature of electronic band states thereby specifying the role of electrons band topology. The presented analysis clarifies the basics of the electron gas magnetoelectric response in two-dimensional magnets and contributes to the ongoing discussion of its spintronics applications.

\section*{Acknoledgments}

The Author thanks I.V.~Rozhansky, M.M.~Glazov, P.S.~Alekseev and N.S.~Averkiev from the Ioffe Institute for a very fruitful discussion of the results and for giving useful advices. 
The work has been carried out with the financial support of the Russian Science Foundation (project 18-72-10111). K.S.D. also thanks the Foundation for the Advancement of Theoretical Physics and Mathematics “BASIS”.

\newpage

\appendix

\section{Wave-packet dynamics semiclassical approach}
\label{ap-1}

The semiclassical theory of band electrons moving in a spatially varying adiabatic perturbation $U(\bm{r})$ can be built by considering the wave-packet dynamics~\cite{xiao2010berry}. 
Let us introduce the wave packet 
$|W_k^n \rangle$ consisting of the $n$-th band Bloch states $|u_k^n \rangle$, its centre of mass coordinates in real and momentum spaces are located at $(\bm{r}_c, \bm{k})$. 
The average of the physical quantity ${Q}$ described by the operator $\hat{Q}$ can be expressed in the following way~\cite{culcer2004semiclassical}
\begin{equation}
\label{eq:A1}
{Q} = \sum_{k, n} f_{n}(\bm{k},\bm{r}) \cdot \langle W_k^{n}| 
\hat{Q} 
| W_k^n \rangle |_{\bm{r} = \bm{r}_c}
- \nabla_{\bm r} \cdot \sum_{k, n} f_{n}(\bm{k},\bm{r}) 
\cdot \langle W_k^{n}| 
\hat{Q} \cdot \left(\hat{\bm{r}} - \bm{r}\right)
| W_k^n \rangle|_{\bm{r} = \bm{r}_c},
\end{equation}
where the first term treats the wave packet as a point particle with the distribution function $f_n(\bm{k},\bm{r})$, and the second term is the first-order correction due to the wave-packet finite size effects. 
The great advantage of this consideration is that it allows one to describe the electron dynamics in terms of semiclassical equations. For instance, in the nondissipative regime $f_n(\bm{k},\bm{r})$ satisfies the Liouville's equation 
\begin{equation}
\frac{d f_n}{dt} = \frac{\partial f_n}{\partial t} + \left\{ 
f_n; \mathcal{H}
\right\} 
= 0, 
\end{equation}
where $\mathcal{H} = \varepsilon_k^n + U(\bm{r})$ is the classical Hamiltonian function in $n$-th electron band with energy~$\varepsilon_k^n$. 
The Poisson bracket $\{A;B\}$ for $A,B$ physical quantities depending on ($\bm{r},\bm{k}$) takes into account the kinematic Berry phase~\cite{shindou2005noncommutative,bliokh2006hamiltonian,duval2006comment}
\begin{align}
& \left\{ 
A; B
\right\} = \omega_{\alpha \beta} \cdot (\partial_\alpha A) (\partial_\beta B), 
\qquad \omega_{\alpha \beta} = \begin{pmatrix}
\varepsilon_{\alpha \beta \gamma} \Omega_\gamma^n & \delta_{\alpha \beta}\\
-\delta_{\alpha \beta} & 0
\end{pmatrix}, 
\qquad
(\alpha,\beta) = ({\bm{r},\bm{k}}), 
\end{align}
where $\omega_{\alpha \beta}$ is the antisymmetric Poisson matrix, $\varepsilon_{\alpha \beta \gamma}$ is the Levi-Civita tensor, and $\bm{\Omega}^n$ is the Berry curvature in $n$-th Bloch band defined as follows
$
\bm{\Omega}^n =
\nabla_{\bm k} \times \mathcal{A}_k^n = 
i \langle 
\nabla_{\bm k} u_{k}^n | \times | \nabla_{\bm k} u_k^n
\rangle, 
$
where $\mathcal{A}_k^n$ is the Berry connection. 
The expression for the Liouville's equation 
with account for the explicit form of $\omega_{\alpha\beta}$ 
is given~by:
\begin{align}
\label{eq:A5}
& 
 \frac{\partial f_n}{\partial t} +
\left(
\frac{\partial \varepsilon_k^n}{\partial \bm{k}} + \left[  \dot{\bm k}  \times \bm{\Omega}_n\right]
\right) \cdot \frac{\partial f_n}{\partial \bm{r}}
+ 
\dot{\bm{k}} 
\cdot 
\frac{\partial f_n}{\partial \bm{k}}
= 0
\end{align}
where $\dot{\bm k} = - \nabla_{\bm r} \mathcal{H} = - \nabla_{\bm r}U(\bm{r})$. The second term in brackets describes a full electron velocity ${\bm v} = \{ \mathcal{H}; \bm{r} \} = \bm{v}_k^n - [\nabla_{\bm r} U, \bm{\Omega}^n]$, 
here $\bm{v}_k^n = \nabla_k \varepsilon_k^n$.

Let us apply this technique to calculate the emerging spin density nearby the electrostatic inhomogeneity. 
We focus on the linear response regime. 
Following Eq.~\ref{eq:A1} we present the spin density $\bm{S}(\bm{r})$ as follows
\begin{align}
\label{eq:A6}
& \bm{S}(\bm{r}) = \sum_{k,n} f_n(\bm{k},\bm{r}) \cdot \langle W_k^{n}| 
\hat{\bm S}
| W_k^n \rangle 
- \nabla_{\bm r} \cdot \sum_{k,n} f_n(\bm{k},\bm{r}) 
\langle u_k^n | \hat{\bm S}  \left(
i \nabla_{\bm k} - \mathcal{A}_k^n
\right)
| u_k^n \rangle. 
\end{align}
In the second term we took into account that the wave packet $| W_k^n \rangle$ is strongly localized nearby $\bm{k}$ in the momentum space and we can approximate it as follows $|W_k^n \rangle \approx e^{i \bm{kr}} | u_k^n \rangle$, which leads us directly to the expression in Eq.~\ref{eq:A6}. 
The unperturbed spin density $\bm{S}_0$ corresponds to $U(\bm{r})=0$, at that $f_n(\bm{k},\bm{r}) = f_n^0(\bm{k})$ and $\bm{S}_0$ is given by
\begin{equation}
\bm{S}_0 = \sum_{k,n} f_n^0(\bm{k}) \langle u_k^n | \hat{\bm S} | u_k^n \rangle.
\end{equation}
The linear order deviations from $\bm{S}_0$ arise from three different origins. Firstly, the distribution function $f_n(\bm{k},\bm{r}) = f_n^0(\bm{k}) + \delta f_n(\bm{k},\bm{r})$ in presence of $U$ is modified according to Eq.~\ref{eq:A5}
\begin{equation}
\left(\bm{v}_k^n \cdot \nabla_{\bm r}\right) \delta f_n(\bm{k},\bm{r}) + \bm{F}(\bm{r}) \cdot \frac{\partial f_n^0}{\partial \bm k} =0, 
\qquad
\delta f_n(\bm{k},\bm{r}) = -U(\bm{r}) \left( - \frac{\partial f_n^0}{\partial \varepsilon}\right).
\end{equation}
Taking into account the redistribution of the electron density in the first term in Eq.~\ref{eq:A6} and approximating $\langle W_k^{n}| 
\hat{\bm S}
| W_k^n \rangle \approx  \langle u_k^n | \hat{\bm S} | u_k^n \rangle$ we obtain the contribution identical with  Eq.~\ref{eq_delta-S1} in the density matrix approach 
\begin{align}
& \delta \bm{S}^{(1)}(\bm{r}) = \sum_{k,n} \delta f_n(\bm{k},\bm{r}) \langle u_k^n | \hat{\bm S} | u_k^n \rangle. 
\end{align}
Also, the inhomogeneous structure of $f_n$ gives rise to the spin-dipole contribution, that is the second term in Eq.~\ref{eq:A6}
\begin{equation}
\label{eq:A10}
\delta \bm{\mathcal{S}}(\bm{r}) = - \bm{F}(\bm{r}) \cdot 
\sum_{k,n} \left( - \frac{\partial f_n^0}{\partial \varepsilon}\right) \langle u_k^n | \hat{\bm S}  \left(
i \nabla_{\bm k} - \mathcal{A}_k^n
\right)
| u_k^n \rangle. 
\end{equation}
The straightforward evaluation of this expression for the Rashba ferromagnet model leads to the susceptibility $\chi_d$ given by Eq.~\ref{eq_chi1d}. 
Finally, there is also the linear order perturbation which is not associated with the change in the electron distribution. In fact, the first term in Eq.~\ref{eq:A6} is determined by the average spin of an electron wave packet $\bm{s}_k^n(t) = \langle W_k^{n}| \hat{\bm S} | W_k^n \rangle $, which 
satisfies the precession equation
\begin{equation}
	\frac{d \bm{s}_k^n}{dt} = \left[\bm{\Omega}_k \times \bm{s}_k^n
	\right]. 
\end{equation}
According to our discussion from~\ref{s_3a}, the wave-packet spin acquires a non-adiabatic correction $\delta \bm{s}_k^n$ linear in $\bm{F}$ and given by Eq.~\ref{eq_s-prec2}.  
This term gives rise to the spin perturbation $\delta \bm{S} = \sum_{(k,n)} f_n^0 \delta \bm{s}_k^n$ identical to $\chi_t$ contribution to the spin susceptibility from Eq.~\ref{eq_chi1d}. 


\section{Kubo formula in the static limit}
\label{ap-2}

In this appendix we relate the semiclassical description of magnetoelectric susceptibility in terms of the density matrix with the Kubo formula for the charge-spin correlation functions, considered in detail in~\cite{denisov2019chiral}. 
The spin density induced in 2DEG by the change in the potential energy $U(\bm{r})$ is given in linear response by
\begin{equation}
\label{eq:B1}
\delta \bm{S}(\bm{r}) = \int \frac{d\bm q}{(2\pi)^2} e^{i \bm{qr}} \bm{\mathcal{Q}}(\bm q) U(\bm{q}), 
\end{equation}
where $U(\bm{q})$ is the Fourier component of $U(\bm{r})$ 
and the static charge-spin correlation function  $\bm{\mathcal{Q}}(\bm q)$ 
can be computed from the Kubo formula
\begin{align}
	\label{eq:B2}
& 
\bm{\mathcal{Q}}(\bm q) = \sum_{m,n} \bm{\mathcal{Q}}^{mn}(\bm q),
\\
&\bm{\mathcal{Q}}^{mn}(\bm q) = \sum_k f_k^m 
\frac{\langle 
	u_k^{m}| \bm{\hat{S}} | u_{k+q}^{n} 
	\rangle
	\langle u_{k+q}^{n} | u_k^{m} \rangle}{\varepsilon_k^{m} - \varepsilon_{k+q}^{n} + i0} 
- f_{k+q}^m 
\frac{\langle 
	u_k^{n}| \bm{\hat{S}} | u_{k+q}^{m} 
	\rangle
	\langle u_{k+q}^{m} | u_k^{n} \rangle}{\varepsilon_k^{n} - \varepsilon_{k+q}^{m} + i0}. 
\notag
\end{align}
The terms $\bm{\mathcal{Q}}^{nn}$ with $m=n$ describe the intraband contributions, while $\bm{\mathcal{Q}}^{mn}$ with $m\neq n$ correspond to the interband ones.

The Kubo formula~\ref{eq:B2} has been explicitly  evaluated for an arbitrary wavevector $\bm{q}$ in~\cite{denisov2019chiral} for Rashba ferromagnet and Dirac models. Here 
we focus on the semiclassical regime when the potential $U$ changes smoothly on the Fermi wavelength $\lambda_F$ scale, so the following relation is fulfilled $ \lambda_F \cdot \nabla_r U \ll U$. 
In this case the spin response becomes local and the correlation function for the Rashba ferromagnet model takes the following form $\bm{\mathcal{Q}} = i \bm{q} \cdot \chi$, where $\chi$ is the $q$-independent coefficient describing the susceptibility $\delta \bm{S}(\bm{r}) = \chi \cdot \bm{E}(\bm{r})$. 

We now proceed with considering the role of intra- and interband terms. 
In the intraband contribution $\bm{\mathcal{Q}}^{nn}$ we replace $(f_k^n - f_{k+q}^n)/(\varepsilon_k^n - \varepsilon_{k+q}^n + i0) \approx \partial f_k^n/\partial \varepsilon$ 
and keep only the ${q}$-linear terms 
in the matrix elements. At that the expression 
takes the following form
\begin{equation}
\bm{\mathcal{Q}}^{nn} 
= -  i \bm{q}  \cdot \sum_k \left( - \frac{\partial f_n^0}{\partial \varepsilon}\right) 
\langle 
 u_k^n |
 \hat{\bm{S}} \left(
 i \nabla_{\bm k} - \mathcal{A}_k^n
 \right)
 |u_k^n  \rangle, 
\end{equation}
where $\mathcal{A}_k^n =  i \langle u_k^n| \nabla_{{k}}  u_k^n \rangle$ is the Berry connection. When taking the Fourier transform Eq.~\ref{eq:B1}  $\bm{\mathcal{Q}}^{nn}$ gives exactly the spin perturbation $\delta \bm{\mathcal{S}}$ in form of Eq.~\ref{eq:A10} corresponding to the spin-dipole term within the semiclassical wave-packet approach. 
We thus conclude that the spin-dipole effect from Eq.~\ref{eq_chi1d} is related to the intraband terms in the Kubo formula. 

In the interband contributions $\bm{\mathcal{Q}}^{mn}$ we also keep only the linear terms with respect to $\bm{q}$, which brings us to the following expression
\begin{align}
& \bm{\mathcal{Q}}^{m n}(\bm{q}) = i \bm{q} \cdot \sum_k f_k^m~
 {\rm Re}\left( 
\frac{
\langle u_k^n | \hat{\bm{\sigma}} | u_k^m \rangle \cdot
{\mathcal{A}}_k^{m n} 
}{\varepsilon_k^m - \varepsilon_k^n}
\right), 
\end{align}
where $\mathcal{A}_k^{m n} = i \langle u_k^m | \nabla_k u_k^n \rangle$. 
The straightforward calculations
for the Rashba ferromagnet model gives
\begin{equation}
 \bm{\mathcal{Q}}^{mn}(\bm{q}) = i \bm{q} \sum_k f_k^m \cdot \frac{\mathcal{F}_k^{mn}}{2\lambda_{so}}, 
\end{equation}
where $\mathcal{F}_k^{mn} = \nabla_{{k}} \times \mathcal{A}_k^{m n} $ is the Berry curvature. The interband terms are related exactly to the non-adiabatic spin tilt effect described by $\delta \bm{s}_k^\eta$ in the density matrix formalism and given by $\chi_t$ susceptibility from Eq.~\ref{eq_chi1d}. 

\section{Kubo formula in the dynamical regime}
\label{ap-3}
In this appendix we relate the Kubo formula calculations of the optical conductivity with the spin resonance related terms emerging in the density matrix approach. 
Kubo formula for the conductivity is given by
\begin{equation}
\sigma_{\alpha \beta}(\omega) = \frac{i e^2}{S} \sum_{k,m,n} \frac{f_{k}^m - f_{k+q}^n}{\varepsilon_{k}^m - \varepsilon_{k+q}^n}\cdot 
\frac{v_{(k,m),(k+q,n)}^\alpha v_{(k+q,n) , (k,m)}^\beta}{\varepsilon_{k}^m - \varepsilon_{k+q}^n + \hbar \omega + i0},
\end{equation}
where $\bm{q} \to 0$ and 
${\bm{v}}_{ij}$ is the proper matrix element of the velocity operator between $i,j$ states. 
We consider firstly the longitudinal conductivity $\sigma_{xx}(\omega)$. The contribution to $\sigma_{xx}(\omega)$ due to intraband terms has the form
\begin{equation}
\sigma_{xx}^{\rm intra}(\omega) = \frac{i e^2}{\hbar \omega} \sum_m \int d\varepsilon~\nu_m(\varepsilon) \left(- \frac{\partial f_k^m}{\partial \varepsilon}\right) \langle | v_{{k},m}^x |^2 \rangle, 
\end{equation}
where $\nu_m$ is the density of states in the corresponding band $m$ 
and $\langle | v_{{k},m}^x |^2 \rangle$ is the angular averaged square of the matrix element modulus.  
This part describes the 
Drude conductivity at $\omega \tau_{sc} \gg 1$ 
due to the perturbation of the electron density 
and it corresponds to Eq.~\ref{eq_long_sigma} from the main text.
For the Rashba ferromagnet model the evaluation of the integral gives 
\begin{equation}
	\label{eq:C3}
	\sigma_{xx}^{\rm intra}(\omega) = i \frac{e^2}{\omega} \cdot \frac{{v_{F+}^2 \nu_F^+} + {v_{F-}^2 \nu_F^-}}{2}. 
\end{equation}

The contribution to $\sigma_{xx}(\omega)$ due to interband terms in case of the Rashba ferromagnet model has the following form 
\begin{equation}
	\sigma_{xx}^{\rm inter}(\omega) = \frac{i e^2}{S} \sum_{k} \frac{f_{k}^- - f_{k}^+}{- \Omega_k} \cdot 
	\frac{
		\langle | v_{(k,-), (k,+)}^x |^2 \rangle
		}{\hbar \omega - \Omega_k + i0} + 
\frac{f_{k}^+ - f_{k}^-}{\Omega_k} \cdot 
\frac{
	\langle | v_{(k,-), (k,+)}^x |^2 \rangle
}{\hbar \omega + \Omega_k + i0}. 
\end{equation}
The angular averaged term is 
$\langle | v_{(k,-), (k,+)}^x |^2 \rangle = (\lambda_{so}^2/2) (1+ \Omega_0^2/\Omega_k^2)$.  
Using this formula and combining the denominators in $\sigma_{xx}^{\rm inter}$ we get the following expression 
\begin{equation}
	\sigma_{xx}^{\rm inter}(\omega) = 
 -i e^2 \cdot 
\sum_k \frac{\left( f_k^- - f_k^+ \right) \hbar \omega}{\left(\hbar \omega - \Omega_k + i \Gamma \right) 
	\left(\hbar \omega + \Omega_k + i \Gamma \right)} \frac{\lambda_{so}^2}{\Omega_k} \left(1 + \frac{\Omega_0^2}{\Omega_k^2}\right), 	
\end{equation}
which repeats Eq.~\ref{eq_sigmal} for $\sigma_l(\omega)$. 
We thus conclude that $	\sigma_{xx}^{\rm inter}(\omega)$ is related to $\left[\bm{n}_k \times \delta \bm{s}_{k0}^\eta \right]$ polarization in terms of the in-plane spin density (see 
Eqs.~\ref{eq_s-dynam},~\ref{eq_static_sko} from the main text).
It is instructive to analyze the energy absorption due to the spin resonance. 
For this purpose we write down explicitly the expression for the real part of the longitudinal conductivity due to the interband terms
\begin{equation}
{\rm Re}[\sigma_{xx}^{\rm inter}(\omega)] = \frac{\pi e^2}{\hbar \omega S} \sum_{k}\left(f_{k}^- - f_{k}^+\right)  \left|  v_{(k,-), (k,+)}^x
\right|^2 \cdot \delta\left(\varepsilon_{k}^- - \varepsilon_k^{+} + \hbar \omega \right).
\end{equation}
The expression has the form of the Fermi golden rule, its straigthforward 
calculation leads to the Eq.~\ref{eq_alpha_abs}. 

We now turn to the transversal component of the conductivity. The interband contribution can be expressed as:
\begin{equation}
\sigma_{yx}^{\rm inter}(\omega) = \frac{i e^2}{S} \sum_k \frac{f_k^- - f_k^+}{- \Omega_k} \frac{v_{-+}^y v_{+-}^x }{\hbar \omega - \Omega_k + i0}
+ 
\frac{f_k^+ - f_k^-}{\Omega_k} \frac{\left(v_{-+}^y v_{+-}^x \right)^\ast}{\hbar \omega + \Omega_k + i0}
\end{equation}
The angular averaged combination of matrix elements $\langle v_{-+}^y v_{+-}^x \rangle = - i \lambda_{so} \Omega_0/\Omega_k$ is purely imaginary. Combining both terms we obtain 
\begin{equation}
\sigma_{yx}^{\rm inter}(\omega) = - \frac{e^2}{S} \sum_{k<k_F^-} \frac{\Omega_k^2}{(\hbar \omega - \Omega_k + i0)(\hbar \omega + \Omega_k + i0)} \cdot \mathcal{F}_k,
\end{equation}
which is same expression Eq.~\ref{eq_chi1_omega} that we get via the density matrix formalism considering $\bm{s}_{\bm{k}0}$ contribution to the spin density (see Eqs.~\ref{eq_s-dynam},~\ref{eq_static_sko} from the main text).

\newpage

\bibliography{Ref}

\end{document}